\documentclass[12pt]{article}
\pdfoutput=1

\usepackage[pdftex]{graphicx}
\usepackage{lscape} 
\usepackage{array}

\usepackage{fix-cm}

\usepackage[makeroom]{cancel}
\usepackage[normalem]{ulem}
\usepackage{amsmath,amssymb}
\usepackage{slashed}
\usepackage{xcolor}
\usepackage{cite}
\usepackage{pdflscape}
\usepackage{multirow}
\usepackage[thinlines]{easytable}
\usepackage{url}
\usepackage[utf8]{inputenc}
\usepackage{longtable}
\usepackage{booktabs}

\newcommand{\eq}[1]{\begin{equation} #1 \end{equation}}
\newcommand{\eqa}[1]{\begin{eqnarray} #1 \end{eqnarray}}

\newcommand{\GeV}{\,{\rm GeV}}
\newcommand{\Eq}[1]{Eq.~(\ref{#1})}
\newcommand{\Ref}[1]{Ref.~\cite{#1}}
\newcommand{\Sec}[1]{Section~\ref{#1}}
\newcommand{\App}[1]{Appendix~\ref{#1}}
\newcommand{\Tab}[1]{Table~\ref{#1}}
\newcommand{\Fig}[1]{Figure~\ref{#1}}
\newcommand{\cL}{{\cal L}}
\newcommand{\K}{{\cal K}}
\newcommand{\cO}{{\cal O}}

\newcommand{\A}{{\cal A}}

\newcommand{\M}{{\cal M}}
\newcommand{\F}{{\cal F}}
\renewcommand{\H}{{\cal H}}
\newcommand{\av}[1]{\langle #1 \rangle}

\begin{document}

%\allowdisplaybreaks

\begin{titlepage}

\vspace*{-2cm}
\begin{flushright}
%TUM-HEP-xxxx/19\\
%MIT-CTP/xxxx\\[5mm]
%{\color{red} Draft of \today}
\end{flushright}

\vspace{2.2cm}

\begin{center}
\bf
\fontsize{19.6}{24}\selectfont
Exact NLO Matching and Analyticity in $b\to s\ell\ell$
\end{center}

\vspace{0cm}

\begin{center}
\renewcommand{\thefootnote}{\fnsymbol{footnote}}
{Hrachia M. Asatrian$^a$, Christoph Greub$^b$ and Javier Virto$^{c}$}
\renewcommand{\thefootnote}{\arabic{footnote}}
\setcounter{footnote}{0}

\vspace*{.8cm}
\centerline{${}^a$\it Yerevan Physics Institute, 0036 Yerevan, Armenia}
\vspace{2.5mm}
\centerline{${}^b$\it Albert Einstein Center for Fundamental Physics, Institute for Theoretical Physics,}
\centerline{\it University of Bern, CH-3012 Bern, Switzerland}
\vspace{1.3mm}
\centerline{${}^c$\it Departament de Física Quàntica i Astrofísica, Institut de Ciències del Cosmos,}
\centerline{\it Universitat de Barcelona, Martí Franquès 1, E08028 Barcelona, Catalunya}\vspace{1.3mm}

\vspace*{.2cm}

\end{center}

\vspace*{10mm}
\begin{abstract}\noindent\normalsize
Exclusive rare decays mediated by $b\to s\ell\ell$ transitions receive contributions from four-quark operators that cannot be naively expressed in terms of local form factors. Instead, one needs to calculate a matrix element of a bilocal operator. In certain kinematic regions, this bilocal operator obeys some type of Operator Product Expansion, with coefficients that can be calculated in perturbation theory. We review the formalism and, focusing on the dominant SM operators $\cO_{1,2}$, we perform an improved calculation of the NLO matching for the leading dimension-three operators. This calculation is performed completely analytically in the two relevant mass scales (charm-quark mass $m_c$ and dilepton squared mass $q^2$), and we pay particular attention to the analytic continuation in the complex $q^2$ plane. This allows for the first time to study the analytic structure of the non-local form factors at NLO, and to calculate the OPE coefficients far below $q^2=0$, say $q^2 \lesssim -10\GeV^2$.
We also provide explicitly the contributions proportional to different charge factors, which obey separate dispersion relations.

\end{abstract}

\end{titlepage}
\newpage 

\renewcommand{\theequation}{\arabic{section}.\arabic{equation}} 
%%%%%%%%%%%%%%%%%%%%%%%%%%%%%%%%%%%%%%%%%%%%%%%

\setcounter{tocdepth}{2}
\tableofcontents

%%%%%%%%%%%%%%%%%%%%%%%%%%%%%%%%% 

\section{Introduction}
\label{sec:intro}
\setcounter{equation}{0}

Exclusive $b\to s\ell\ell$ decays such as $B\to K^{(\star)}\ell\ell$ and $B_s\to \phi\ell\ell$ have been on the focus point of theorists and experimentalists for some time, due to the potential they provide for tests of the Standard Model (SM).
While the interest for such decays dates back to the era of the $B$-factories (which provided some of the first measurements), a renewed interest has been triggered by the measurements at the LHC, most prominently the ones by the LHCb collaboration.
Starting with the ``$P'_5$ Anomaly"~\cite{Descotes-Genon:2013wba,Aaij:2013qta}, and followed by a larger pattern of ``tensions" of different degrees in the landscape of angular and dilepton-mass-squared distributions in $B_{(s)}\to \{K^{(\star)},\phi\}\mu^+\mu^-$ modes~\cite{Descotes-Genon:2015uva,Altmannshofer:2014rta}, these measurements (in inseparable association with theoretical work) have guided the community during the LHC era. More precise experimental studies are part of the programs for the LHC upgrade~\cite{LHCb:2008aa} and Belle-II~\cite{Kou:2018nap}, and there is little doubt they will lead to new discoveries. The question is whether these discoveries will involve Beyond-the-SM (BSM) or QCD/hadronic physics. While this is subject to the personal inclination of the reader, both outcomes are truly interesting.

The exclusive $b\to s\ell\ell$ decays belong to the class of ``rare" FCNC transitions which are loop-, CKM- and GIM-suppressed in the SM. This leads to branching fractions of the order of~$10^{-6}$, and which could be easily altered by BSM physics lifting any of such suppression mechanisms. However, it is increasingly evident that large deviations with respect to the SM are not present, and as such, rare decays are no longer smoking guns of BSM physics. Thus we need to test SM predictions more precisely. This is now possible due to the large statistics collected at the LHC (with more than 2K selected $B\to K^{\star}\mu\mu$ events in Run~1 by LHCb),
but it also implies that theory predictions with uncertainties below $\sim 10\%$ are necessary, with model dependence reduced to the minimum.

Theory predictions for $B \to M\ell^+\ell^-$ observables depend on non-perturbative hadronic matrix elements of two types: ``local" and ``non-local" form factors~(e.g.~\cite{Bobeth:2017vxj}). Contributions to the amplitude from semileptonic ($[\bar s \Gamma b][\bar \ell \Gamma' \ell]$) or dipole ($[\bar s \sigma_{\mu\nu} P_R b] F^{\mu\nu}$) operators are exactly factorizable and proportional to local form factors --matrix elements of local fermionic currents-- to all orders in QCD (but to the leading order in QED effects).
These local form factors are known relatively well and can be calculated with Light-Cone Sum Rules \mbox{(LCSRs)} or Lattice~QCD (LQCD) methods, both agreeing well with each other~\cite{Lattice:2015tia,Bouchard:2013pna,Horgan:2013hoa,Straub:2015ica,Gubernari:2018wyi,Descotes-Genon:2019bud}.
On the contrary, contributions from four-quark operators such as $[\bar s \gamma_\mu P_L c][\bar c \gamma^\mu P_L b]$ are proportional to non-local form factors, more precisely, the matrix elements of time-ordered products of a four-quark operator and an electromagnetic current. 
The calculation of these non-local form factors is highly non-trivial and relies inevitably on some type of operator-product expansion (OPE)~\cite{Khodjamirian:2010vf,Grinstein:2004vb,Beylich:2011aq}.
In this way, the complicated non-local form factors can be written in terms of simpler hadronic matrix elements, multiplied by coefficients that can be determined though a perturbative matching calculation.
These simpler hadronic matrix elements are either local form factors, or matrix elements of bi-local operators defined on the light-cone, which can be expressed in terms of meson light-cone distribution amplitudes.

The matching to the leading (dimension-three) operators in the OPE can be extracted from the perturbative partonic calculation of the matrix element, which has been known up to order $\alpha_s$ (two loops) for some time~\cite{Asatryan:2001zw,Ghinculov:2003qd,Greub:2008cy}, 
albeit not in full analytic form in the two relevant variables: $q^2$ (the dilepton squared invariant mass) and $m_c$ (the charm-quark mass). Only recently the necessary analytic calculation of the two-loop master integrals involved has been achieved~\cite{Bell:2014zya}, and applied to the problem at hand~\cite{deBoer:2017way}.

We have repeated the full analytic two-loop calculation independently, and checked the results of~\Ref{Bell:2014zya}, which we confirm. The explicit and independent check of this calculation is the first result of this paper. But we have done the calculation in a way that lays out the analytic structure of the results more explicitly, and imposing an analytic continuation which is more convenient for the dispersive analysis (see Refs.~\cite{Bobeth:2017vxj,Khodjamirian:2012rm}).
%This requires fixing the boundary conditions in the differential equations for the master integrals at two different points in the complex $q^2$ plane.
The results in this form allow us to study the branch cut discontinuities and compare them with the expectations derived from unitarity, as well as to test \emph{all} the analytic singularities of the two-loop amplitude by explicitly checking a dispersion relation. 
This is the second result of this paper.
Finally, the dispersion relation formalism is an important tool to extend consistently the calculations in the LCOPE region (negative $q^2$) to the physical region at $q^2> 0$. Under certain simplifying assumptions, this dispersion relation can be separated in pieces multiplying difference quark charge factors. For that purpose the NLO contributions to the OPE coefficients must also be separated in this way, but this separation has not yet been given explicitly. We do give separate contributions to the OPE coefficients to be used in the separated dispersion relations, which is the third result of this paper.

We start in~\Sec{sec:framework} by reviewing the theoretical framework and fixing the conventions and the notation.
In~\Sec{sec:calculation} we give the details of the analytic NLO matching calculation.
In~\Sec{sec:numerics} we address the issue of the numerical evaluation of the NLO functions, which requires some care due to the presence of Generalized Polylogarithms (GPLs) up to weight four. We also compare our results with the ones in the literature, and provide explicit numerical results at various kinematic points in the LCOPE region.
In~\Sec{sec:analytic} we discuss the analytic properties of the results and prove the structure of singularities by means of a dispersion relation. We then explain how to separate the NLO matching coefficients into the two contributions proportional to different charge factors.
We conclude in~\Sec{sec:conclusions}.
The various appendices include supplementary information on: A. The attached ancillary files which contain all our results in electronic form as well as codes for numerical evaluations; B. The list of the relevant Master Integrals that appear in the calculation of the two-loop diagrams; C. The list of different weights appearing in the GPLs in the results; and D. A few examples on fixing the integration constants that arise in the calculation of the two-loop Master Integrals.

\section{Theoretical framework}
\label{sec:framework}
\setcounter{equation}{0}

\subsection{Set-up: Weak Effective Theory and Conventions}

$B$ decay amplitudes are calculated within the Weak Effective Theory (WET) where the SM particles with EW-scale masses have been integrated out. The WET lagrangian then contains QCD and QED interactions, and a tower of higher dimensional local operators which is typically truncated at dimension six~\cite{Buchalla:1995vs,Aebischer:2017gaw}. The part of the WET Lagrangian which is relevant for the contributions discussed in this paper is:
\eq{
\cL_{\rm WET} = \cL_{\rm QCD} + \cL_{\rm QED} + \frac{4G_F}{\sqrt2} V_{ts}^*V_{tb} \Big[
C_1 \cO_1 + C_2 \cO_2  + C_7 \cO_7 + C_9 \cO_9 + C_{10} \cO_{10}
\Big]
\label{LWET}
}
where
\begin{align}
&\cO_1 = (\bar s \gamma_\mu P_L T^a c)(\bar c \gamma^\mu P_L T^a b)\ ,
&&\cO_2 = (\bar s \gamma_\mu P_L c)(\bar c \gamma^\mu P_L b) \ ,
\nonumber\\[2mm]
&\cO_9 = \frac{\alpha}{4\pi} (\bar s \gamma_\mu P_L b)(\bar \ell \gamma^\mu \ell)\ ,
&& \cO_7 = \frac{e}{(4\pi)^2} m_b (\bar s \sigma_{\mu\nu} P_R b) F^{\mu\nu}\ ,
\\[0mm]
&\cO_{10} = \frac{\alpha}{4\pi} (\bar s \gamma_\mu P_L b)(\bar \ell \gamma^\mu \gamma_5 \ell)\ ,
&&
\nonumber
\end{align}
We use the following conventions:
$P_{R,L}=(1\pm\gamma_5)/2$,
$\sigma_{\mu\nu}\equiv (i/2) [\gamma_\mu,\gamma_\nu]$,
the covariant derivative is given by $D_\mu q = (\partial_\mu + i e Q_q A_\mu + i g_s T^A G^A_\mu)q$, and $m_b=m_b(\mu)$ denotes the $\overline {\rm MS}$ $b$-quark mass. In our calculation of NLO corrections from $\cO_{1,2}$, the scheme dependence of $m_b$ is a higher order effect. We will neglect the strange quark mass throughout the paper.

\subsection{Local and Non-local form factors in exclusive $b\to s\ell^+\ell^-$}

To the leading non-trivial order in QED, the effective theory amplitude for the exclusive decay $\bar B\to M \ell^+\ell^-$, with $M$ an undetermined meson (or hadronic state in general~\cite{Descotes-Genon:2019bud}), is given in terms of local and non-local form factors~\cite{Bobeth:2017vxj,Beneke:2001at,Descotes-Genon:2019bud}:
\eq{
\A(\bar B\to M \ell^+\ell^-) =  \frac{G_F\, \alpha\, V^*_{ts} V_{tb}}{\sqrt{2} \pi} \bigg[ (C_9 \,L^\mu_{V} + C_{10} \,L^\mu_{A})\  \F_\mu 
-  \frac{L^\mu_{V}}{q^2} \Big\{  2 i m_b C_7\,\F^{T}_\mu  + \H_\mu \Big\}   \bigg]\ ,
\label{AmplitudeBMll}
}
up to terms of $\cO(\alpha^2)$.
Here $q^2$ is the invariant squared mass of the lepton pair and
$L_{i}^\mu$ are leptonic currents, $L_{V(A)}^\mu \equiv \bar u_\ell(q_1) \gamma^\mu(\gamma_5) v_\ell(q_2)$.
In this amplitude we have neglected contributions from other local semileptonic and dipole operators that are not relevant in the SM, as well as higher order QED corrections, but it is exact in QCD. All non-perturbative effects are contained in the ``local" and ``non-local" form factors $\F_i^{(T)\mu}$ and $\H^\mu$, with
\eq{
\F_\mu =  \langle M(k)|\bar{s}\gamma_\mu P_L\, b|\bar{B}(q+k)\rangle
\ ,\quad
\F^{T}_\mu = \langle M(k)|\bar{s}\sigma_{\mu\nu} q^\nu P_R\, b|\bar{B}(q+k)\rangle
\ .
}
This paper deals with the non-local form factors $\H^\mu(q,k)$, defined by the following matrix element:
\eq{
\H^\mu(q,k) = 16\pi^2\,i\!  \int d^4x\, e^{i q\cdot x}\,
\av{M(k) | T\big\{ j^\mu_{\rm em}(x), (C_1\cO_1 + C_2\cO_2)(0) \big\}  | \bar B(q+k)}\ ,
\label{eq:H}
}
where $j^\mu_{\rm em} = \sum_{q} Q_q\ \bar q\gamma^\mu q$, with $q=\{u,d,s,c,b\}$. This corresponds to the matrix element of the non-local operator:
\eq{
\K^\mu(q) = 16\pi^2\,i\! \int d^4x\, e^{i q\cdot x}\,
 T\big\{ j^\mu_{\rm em}(x), (C_1\cO_1 + C_2\cO_2)(0) \big\}
\ ,
\label{eq:K}
}
which is the focus of the following discussion.

\subsection{Operator Product Expansion for Non-local form factors}

A reliable calculation of $\H^\mu(q,k)$ is very important for phenomenology and a challenge for theory. At low hadronic recoil, $q^2\sim m_b^2$, the $dx$ integral in~\Eq{eq:K} is dominated by the region $x\sim 1/m_b$, and a local OPE exists for the operator $\K^\mu(q)$~\cite{Grinstein:2004vb,Beylich:2011aq}:
\eq{
\K^\mu_{\rm OPE}(q) =
\Delta C_9(q^2) \big( q^\mu q^\nu - q^2 g^{\mu\nu} \big)
\,\bar s \gamma_\nu P_L b
+
\Delta C_7(q^2)\, 2im_b \,\bar s \sigma^{\mu\nu}q_\nu P_R b
+
\cdots
\label{eq:KOPE}
}
where we have indicated the contribution of operators of dimension three (according to the counting in~\Ref{Beylich:2011aq}), and the ellipsis denotes contributions of operators of higher dimension $d>3$, with OPE coefficients that are suppressed by $m_b^{3-d}\sim (\sqrt{q^2})^{3-d}$. This equation defines the OPE coefficients $\Delta C_{7,9}$.

At large hadronic recoil, and below the on-shell branch cuts, $q^2 \lesssim 0$, the $dx$ integral in~\Eq{eq:K} is instead dominated by the region\footnote{
Here $m_q$ refers to the mass of the quark responsible for the partonic $q\bar q$ branch cut in the variable $q^2$.
} $x^2\sim 1/(4m_q^2 - q^2)$, which allows for a light-cone OPE (LCOPE), where local operators with an arbitrary number of covariant derivatives along the relevant light-cone direction contribute at the same order~\cite{Khodjamirian:2010vf}. 
The structure of the LCOPE coincides with the local OPE at dimension three, and therefore~\Eq{eq:KOPE} is also true at $q^2\lesssim 0$. The power corrections are, however, different. Power corrections to both OPE expansions have been discussed in e.g.~Refs.~\cite{Khodjamirian:2010vf,Khodjamirian:2012rm,Beylich:2011aq}.

Given~\Eq{eq:KOPE}, the non-local form factors~(\ref{eq:H}) are determined by the OPE coefficients and the local form factors:
\eq{
\H^\mu_{\rm OPE}(q^2) = \Delta C_9(q^2) \big(  q^\mu q^\nu - q^2 g^{\mu\nu} \big) \F_\nu
+ 2im_b \,\Delta C_7(q^2) \F^{T\mu} + \cdots \ ,
\label{eq:HOPE}
}
with the ellipsis denoting contributions from subleading terms in the (LC)OPE.
Thus, the effect of the non-local contribution $\H_\mu$ in the amplitude~(\ref{AmplitudeBMll}) at this order in the OPE expansion can be absorbed into ``effective" Wilson coefficients
$C_{7,9}^{\rm eff}(q^2) = C_{7,9} + \Delta C_{7,9}(q^2)$. These effective Wilson coefficients are scheme and scale independent.
The same structure arises to all orders in QCD in the ``factorization approximation", where all interactions between the charm loop and the constituents of the external mesons are neglected. However the OPE formalism beyond the leading order includes all non-factorizable contributions, which appear to be phenomenologically very relevant~\cite{Lyon:2014hpa}.

\subsection{Structure of the OPE matching calculation}

The OPE coefficients $\Delta C_{7,9}(q^2)$ are calculable order by order in perturbation theory through a matching calculation. The easiest way to perform this matching is to equate the matrix elements of partonic states at each order in $\alpha_s$:
\eq{
\M^\mu(q) \equiv \av{s(k)| \K^\mu(q) |b(q+k)} \stackrel{!}{=}
\av{s(k)| \K_{\rm OPE}^\mu(q) |b(q+k)}\equiv \M_{\rm OPE}^\mu(q)\ .
\label{eq:matching}
}
We shall refer to the matrix element $\M^\mu(q)$ in the left-hand side as the ``QCD amplitude" and the one in the right-hand side $\M_{\rm OPE}^\mu(q)$ as the ``OPE amplitude". 
A perturbative calculation of the QCD amplitude leads to an expression of the form:
\eqa{
\M^\mu(q) &=& 
f^{(9)}(q^2)\,
\Big( q^\mu q^\nu - q^2 g^{\mu\nu} \Big)\,
\bar u_s \gamma_\nu P_L u_b
+ f^{(7)}(q^2)\,2im_b\,\bar u_s \sigma^{\mu\nu} q_\nu P_R u_b
\ ,\qquad
\label{eq:f79def}
}
which defines the functions $f^{(7,9)}(q^2)$.
At the leading order (after renormalization),
\eqa{
f_{{\rm LO}}^{(7)}(q^2)&=&0\ ,
\nonumber\\[3mm]
f_{{\rm LO}}^{(9)}(q^2)
&=& \frac{2Q_c (C_F C_1 + C_2)}{3}\, \bigg\{ \frac23 + i\pi + \frac{4z}{s} + \log{\frac{4\mu^2}{m_b^2}} + 2\log{x} - \log{(1-x)} - \log{(1+x)}
\nonumber\\
&& + \frac{1-3y^2}{2y^3} \Big[  \log{(1+y)} - \log{(1-y)} \Big] \bigg\}\ .
\label{eq:f9LO}
}
Here we have defined
\eq{
z = \frac{m_c^2}{m_b^2}\ ,\quad
s = \frac{q^2}{m_b^2}\ ,\quad
x=\frac1{\sqrt{1-4z}}\ ,\quad
y=\frac1{\sqrt{1-4z/s}}\ .
}
The same calculation for the OPE side in~\Eq{eq:matching} is written as:
\eqa{
\M_{\rm OPE}^\mu(q) &=&
h^{(9)}(q^2) \Delta C_9(q^2) \Big( q^\mu q^\nu - q^2 g^{\mu\nu} \Big)\,
\bar u_s \gamma_\nu P_L u_b
+ h^{(7)}(q^2)\, \Delta C_7(q^2)\,2im_b\,\bar u_s \sigma^{\mu\nu} q_\nu P_R u_b
\nonumber\\
&& +\ \text{next order in the OPE expansion} \ ,
}
where, to leading order,
\eq{
h^{(9)}_{\rm LO}(q^2) =  h^{(7)}_{\rm LO}(q^2) = 1 \ .
}
Thus, the leading order matching gives
\eq{
\Delta C_7(q^2) = \cO(\alpha_s)\ ;\quad
\Delta C_9(q^2) = f^{(9)}_{\rm LO}(q^2)  + \cO(\alpha_s) \ .
\label{eq:LOMatchingC9}
}
Beyond the leading order, we write,
\eqa{
f^{(7,9)}(q^2) &=& f_{\rm LO}^{(7,9)}(q^2)
+ \frac{\alpha_s}{4\pi} f_{\rm NLO}^{(7,9)}(q^2) + \cdots\ ,
\\[2mm]
h^{(7,9)}(q^2) &=& h_{{\rm LO}}^{(7,9)}(q^2)
+ \frac{\alpha_s}{4\pi} h_{{\rm NLO}}^{(7,9)}(q^2) + \cdots\ ,
}
which leads to the following NLO matching equations,
\eqa{
\Delta C_7(q^2) &=& \frac{\alpha_s}{4\pi} f^{(7)}_{\rm NLO}(q^2) + \cO(\alpha_s^2)\ ,
\label{eq:NLOMatchingC7}
\\[3mm]
\Delta C_9(q^2) &=& 
f^{(9)}_{\rm LO}(q^2) + \frac{\alpha_s}{4\pi} \Big[
f^{(9)}_{\rm NLO}(q^2) - f^{(9)}_{\rm LO}(q^2)\,h^{(9)}_{\rm NLO}(q^2)
\Big]+ \cO(\alpha_s^2)
 \ .
\label{eq:NLOMatchingC9}
}
As it should be, these coefficients are infrared-finite.
In particular, while $f^{(9)}_{\rm NLO}$ and $h^{(9)}_{\rm NLO}$ are separately infrared-divergent, the divergence cancels in the difference.
The various prefactors in the definition of $f^{(7,9)}$ in~\Eq{eq:f79def} have been chosen such that the contribution from $\cO_{1,2}$ to the $b\to s\ell\ell$ partonic amplitude is
\eq{
\av{s\ell\ell |C_1\cO_1+C_1\cO_2|b} =
f^{(9)}(q^2)\, \av{\cO_9}_{\rm tree} + f^{(7)}(q^2)\, \av{\cO_7}_{\rm tree}
\label{eq:<CiOi>}
}
to all orders in QCD. This makes contact with the notation of~\Ref{Asatryan:2001zw},
\eqa{
f_{\rm NLO}^{(7)}(q^2) &=& -C_1 F_1^{(7)}(q^2) - C_2 F_2^{(7)}(q^2) \ ,
\nonumber\\[2mm]
f_{\rm NLO}^{(9)}(q^2) - f^{(9)}_{\rm LO}(q^2)\,h_{\rm NLO}^{(9)}(q^2) &=& -C_1 F_1^{(9)}(q^2)-C_2 F_2^{(9)}(q^2) \ .
\label{eq:Fi79def}
}
In~\Ref{Asatryan:2001zw} the functions $F_i^{(7,9)}(q^2)$ were calculated at low $q^2$ and the results were represented as expansions in the small parameters $q^2/m_b^2$, $z\equiv m_c^2/m_b^2$ and $q^2/(4 m_c^2)$.
In~\Ref{Greub:2008cy} the functions $F_i^{(7,9)}(q^2)$ were calculated for the high $q^2$ range and the results were given as an expansion in $z$.
In~\Sec{sec:calculation} we describe the calculation of these NLO functions $F_i^{(7,9)}(q^2)$ in a fully analytic form for $z$ and~$q^2$. The full results are discussed in~\Sec{sec:NLOresults}.

\subsection{Analytic structure and dispersion relations}
\label{sec:AnalStr}

In order to discuss the analytic structure of the non-local form factors, it is convenient to perform a Lorentz decomposition and focus on invariant functions:
\eq{
\H^\mu(q,k) = \sum_\lambda \H_\lambda (q^2) \,\eta_\lambda^\mu
}
where $\eta_\lambda^\mu$ are a set of orthogonal Lorentz vectors depending on $q$ and $k$ and $\H_\lambda (q^2)$ are a set of invariant non-local form factors (see e.g.~\Ref{Bobeth:2017vxj}).

Once the non-local matrix elements $\H_\lambda(q^2)$ are known in the OPE regions of the $q^2$ plane, it remains to use this information to extrapolate the results to the physical regions of interest, within the range $0\le q^2 \le (M_B-M_M)^2$.
For this we need some information about the properties of the functions $\H_\lambda(q^2)$ in the complex $q^2$ plane. The most important of such properties is the analytic structure (the structure of their analytic singularities), that is, the presence of poles and branch cuts. Assuming the principle of maximum analyticity, these singularities are fully determined by the on-shell cuts of the matrix elements (see e.g.~\Ref{Bobeth:2017vxj}).

The first thing to note is that, independently of the value of $q^2$, the functions $\H_\lambda(q^2)$ are complex-valued due to on-shell intermediate states in the $p^2$ channel, e.g. $B\to \overline D D_s \to M_\lambda\,\gamma^*$.
The singularity structure associated with the variable $q^2$ will then apply separately to the real and imaginary parts of $\H_\lambda(q^2)$: $\H_\lambda^{\rm (re)}(q^2)$ and  $\H_\lambda^{\rm (im)}(q^2)$. Each of these two functions are then real for $q^2<0$, but develop imaginary parts due to on-shell states in the $q^2$ channel, for $q^2>0$. All these on-shell states must have the (QCD-conserved) quantum numbers of the e.m.~current, which means that (in full QCD) they are necessarily multiparticle states. Therefore the singularities are branch cuts, one for each multiparticle state: $B\to M_\lambda X^{1--}\to M_\lambda  \gamma^*$, 
with $X^{1--} =\{ \pi\pi, \pi\pi\pi, \overline K K, \cdots, \overline D D, \overline D D^*, \cdots\}$. Each of these branch cuts starts at its corresponding threshold $s_{\rm th} = \{ 4m_\pi^2, 9m_\pi^2, 4 m_K^2, \cdots, 4m_D^2, (m_D+m_{D^*})^2, \cdots \}$.

Given the analytic structure of the functions $\H_\lambda(q^2)$, one can write a dispersion relation to relate the values of these functions at specific points to an integral over the branch-cut discontinuity~\cite{Khodjamirian:2012rm}:
\eq{
\H_\lambda(q^2) = \H_\lambda(q_0^2) +
(q^2 - q^2_0) \int_{s_{\rm th}}^\infty dt\,
\frac{\rho_\lambda(t)}{(t-q^2-i\epsilon)(t-q^2_0)}\ ,
\label{eq:disprel}}
where
\eq{
\rho_\lambda(t) = \frac{\H_\lambda(t+i\epsilon)-\H_\lambda(t-i\epsilon)}{2\pi i}
}
is the discontinuity along the cut (the spectral function).
The spectral function $\rho_\lambda(t)$ may, in certain approximations, contain poles below the multiparticle threshold, and thus in such cases the parameter $s_{\rm th}$ is  assumed to lie below such poles.
The subtraction at $q^2_0$ is implemented to ensure the convergence of the dispersion integral~\cite{Khodjamirian:2012rm}. While this dispersion relation is completely general, we assume that $q^2_0$ is within the OPE region (thus $\H_\lambda(q_0^2) = \H_\lambda^{\rm OPE}(q_0^2)$), and $q^2$ can be on the physical range, and thus the $i\epsilon$ prescription in the denominator is chosen such that for (real) $q^2>s_{\rm th}$, the pole in the integrand is above the real axis.
This prescription can be ignored if $q^2$ is away from the branch cut.

\bigskip

One can now separate the different contributions to the e.m.~current in~\Eq{eq:H}, and write three different dispersion relations for $\H_{\lambda,sb}$, $\H_{\lambda,c}$ and $\H_{\lambda,ud}$~\cite{Khodjamirian:2012rm}. These three dispersion relations are equivalent to~\Eq{eq:disprel}, but with two qualifications: (1) the spectral densities also depend on the channel, $\rho_{\lambda,sb}$, $\rho_{\lambda,c}$ and $\rho_{\lambda,ud}$, and (2) the OPE functions $\H_{\lambda,x}^{\rm OPE}(q_0^2)$ correspond to the terms in $\H_{\lambda}^{\rm OPE}(q_0^2)$ proportional to $Q_{s/b}$, $Q_c$, $Q_{u/d}$ for $x=sb,c,ud$. The reason that the terms with $Q_s$ and $Q_b$ are not separated is because they are not separately gauge invariant (see \Sec{sec:IBP}), while the terms with $Q_u$ and $Q_d$ do not receive contributions from the two-loop matching corrections discussed in this paper, and will also depend on the charge of the decaying $B$ meson. The explicit separation into terms with different charge factors $Q_{s/b}$ and $Q_c$ is one of the results in this paper that was not available before. The two-loop contributions to $\H_{\lambda,sb}^{\rm OPE}(q_0^2)$ and $\H_{\lambda,c}^{\rm OPE}(q_0^2)$ will come respectively from diagrams $\{a,b\}$, and $\{c,d,e\}$ in~\Fig{fig:diags}.
Other contributions from CKM-suppressed operators (with $u,d,s$ loops) will contribute to $\H_{\lambda,ud}^{\rm OPE}(q_0^2)$ and $\H_{\lambda,sb}^{\rm OPE}(q_0^2)$. These corrections are simpler than the ones discussed in this paper (since they contain one fewer mass scale) and can be found in analytical form elsewhere~\cite{Seidel:2004jh}.

Up to this point the discussion is rigorous and exact, relying only on maximum analyticity and unitarity.
The separation into different charge factors has been performed to implement a simplifying assumption when modelling the spectral densities, based on OZI suppression~\cite{Khodjamirian:2012rm,Bobeth:2017vxj}.
Up to OZI-suppressed effects, the QCD spectral densities $\rho_{\lambda,sb}$, $\rho_{\lambda,c}$ and $\rho_{\lambda,ud}$ receive separable contributions from intermediate states $\{\phi, K\bar K, \dots\}$, $\{J/\psi,\psi(2S),D\bar D \dots\}$ and $\{\rho,\omega,\pi\pi, \dots\}$, respectively~\cite{Khodjamirian:2012rm,Bobeth:2017vxj}. Therefore the dispersion relation can be divided into three separate ones~\cite{Khodjamirian:2012rm}:
\eq{
\H_{\lambda,x}(q^2) = \H^{\rm OPE}_{\lambda,x}(q_0^2) +
(q^2 - q^2_0) \int_{s_{\rm th}}^\infty dt\,
\frac{\rho_{\lambda,x}(t)}{(t-q^2-i\epsilon)(t-q^2_0)}\ ,
\label{eq:3DRs}
}
with $x=\{c,sb,ud\}$, and 
\eqa{
\rho_{\lambda,c}(t) &=& \frac23 f_{J/\psi}\, \A_\lambda^{J/\psi} \,\delta(t-M_{J/\psi}^2) + \frac23 f_{\psi(2S)}\, \A_\lambda^{\psi(2S)} \,\delta(t-M_{\psi(2S)}^2) + \cdots \ ,
\\
\rho_{\lambda,sb}(t) &=& -\frac13 f_{\phi}\, \A_\lambda^{\phi} \,\delta(t-M_{\phi}^2) + \cdots\ ,
\\
\rho_{\lambda,ud}(t) &=& \frac1{\sqrt2} f_{\rho}\, \A_\lambda^{\rho} \,\delta(t-M_{\rho}^2) + \frac1{3\sqrt2} f_{\omega}\, \A_\lambda^{\omega} \,\delta(t-M_{\omega}^2) + \cdots\ .
}
For consistency with the adopted approximation we have assumed that the resonances below the multi-particle thresholds in each channel are stable, and indicated only these poles in the spectral densities. The ellipses denote the subsequent continuum contributions with open flavors (e.g. ${D\bar D, D^*\bar D, \cdots}$ in $\rho_{\lambda,c}(t)$). The flavor separation of the dispersion relations has some phenomenological advantages~\cite{Khodjamirian:2012rm,Bobeth:2017vxj}.

\section{OPE matching calculation at NLO}
\label{sec:calculation}
\setcounter{equation}{0}

\subsection{OPE functions at NLO and cancellation of IR divergencies}

\begin{figure}
\begin{center}
\includegraphics[width=15cm]{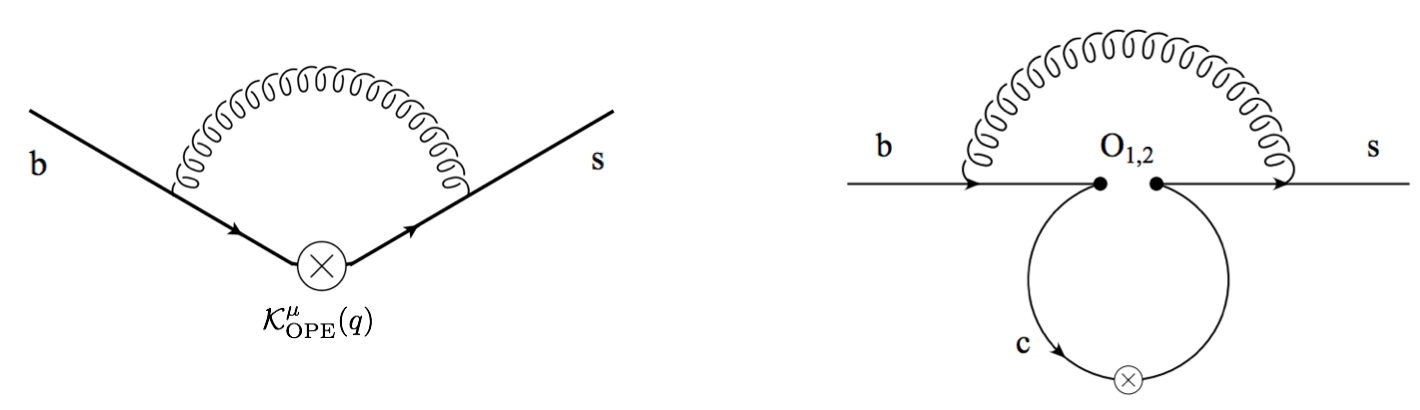}
\end{center}
\caption{
Left: Contribution to the OPE function $h^{(7,9)}_{\rm NLO}$. Since $\Delta C_7 = \cO(\alpha_s)$, $h^{(7)}_{\rm NLO}$ does not contribute to the NLO matching.
Right: Contribution to $f^{(9)}_{\rm NLO}$ which is equal to $f^{(9)}_{\rm LO}\,h^{(9)}_{\rm NLO}$. The contribution of this diagram to $f^{(7)}_{\rm NLO}$ vanishes (since $f^{(7)}_{\rm LO}=0$).
}
\label{fig:diagsIR}
\end{figure}

The NLO functions $h^{(7,9)}_{\rm NLO}$ arise from the diagram in~\Fig{fig:diagsIR} (left). According to the matching equations~(\ref{eq:NLOMatchingC7}),\,(\ref{eq:NLOMatchingC9}), only $h^{(9)}_{\rm NLO}$ is needed for the NLO matching. On the other hand, the contribution to the function $f^{(9)}_{\rm NLO}$ given in~\Fig{fig:diagsIR} (right) is equal to $f^{(9)}_{\rm LO}\,h^{(9)}_{\rm NLO}$, since the LO matching expression $\Delta C_{9,{\rm LO}} = f^{(9)}_{\rm LO}$ ensures that the charm loop can be replaced by $\K^\mu_{\rm OPE}$ at this order in the perturbative expansion. Thus, the two contributions will cancel in the combination
$\big[f^{(9)}_{\rm NLO}(q^2) - f^{(9)}_{\rm LO}(q^2)\,h^{(9)}_{\rm NLO}(q^2)\big]$ in~\Eq{eq:NLOMatchingC9}.
This cancellation is important because these are the only two contributions which are IR divergent.
As a result, the NLO contributions in~\Eq{eq:Fi79def} are obtained by evaluating the five classes of diagrams in~\Fig{fig:diags}.

\subsection{Two loop contributions to the QCD amplitude}
\label{sec:3.2}

\begin{figure}
\includegraphics[width=\textwidth]{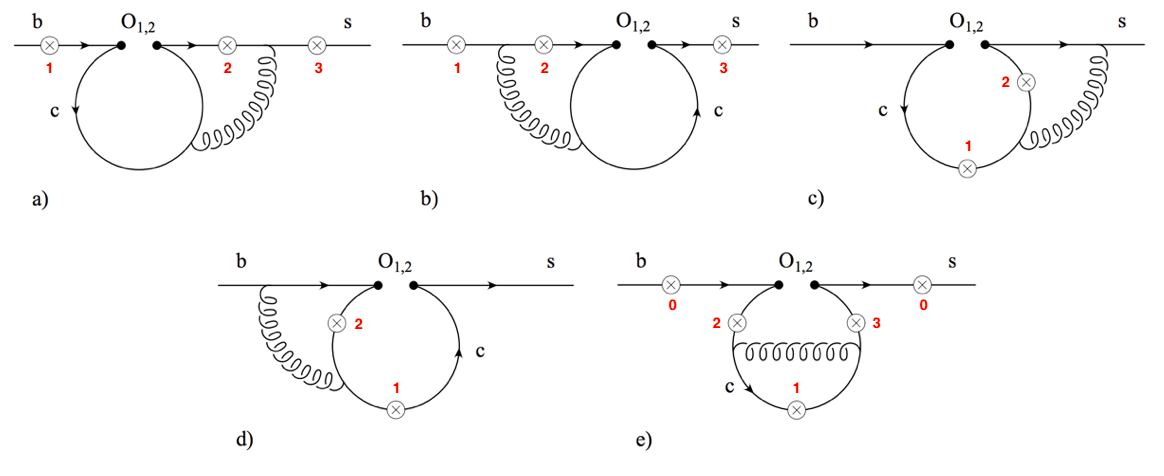}
\caption{The five classes of two-loop diagrams that contribute to the functions $F_{1,2}^{(7,9)}$.
Crosses denote insertion of the EM current, which are numbered for proper reference.
The two diagrams of type (e) labeled as `0' vanish.}
\label{fig:diags}
\end{figure}

The contribution to the QCD amplitude from any given set of Feynman diagrams in~\Fig{fig:diags} can be written as 
\eq{
\av{s(k)| \K^\mu(q) |b(q+k)}|_{{\rm diagrams}\,(i)} =  \bar u_s (p-q) P_R V_{(i)}^\mu(q^2) u_b(p)\ .
}
Conservation of the e.m. current implies that $V_{(i)}^\mu$ has the structure of~\Eq{eq:f79def}:
\eq{
V_{(i)}^\mu(q^2)=
\frac{1}{16 \pi^2} \bigg\{
f_{(i)}^{(9)}(q^2)\,
\big(  q^\mu q^\nu - q^2 g^{\mu\nu} \big)\,\gamma_\nu
+ 2 f_{(i)}^{(7)}(q^2)\,im_b\,\sigma^{\mu\nu} q_\nu \bigg\}
\ ,
}
which is a consequence of the Ward Identity to be checked from the calculation.
In the calculation of $V_{(i)}^\mu$, we use the EOM for the quark spinors
(keeping $m_b\ne 0$ but setting $m_s=0$ here) to remove all factors of  $\slashed p$ and $\slashed q$, and we set $p^2=m_b^2$ and
$(p-q)^2=m_s^2\to 0$. At the end one finds that $V_{(i)}^\mu$ has the form:
\eq{
V_{(i)}^\mu(q^2) = A_{(i)}\, q^\mu + B_{(i)}\, p^\mu + C_{(i)}\, \gamma^\mu
}
where $A_{(i)}$, $B_{(i)}$ and $C_{(i)}$ are scalar functions of $m_b$, $m_c$ and $q^2$. On dimensional grounds, $A_{(i)},B_{(i)}\sim m$ and $C_{(i)}\sim m^2$.
From these coefficients one can read off the functions $f_{(i)}^{(7,9)}(q^2)$ and check the Ward Identity. From $A_{(i)}$ and $B_{(i)}$ one has:
\eq{
f_{(i)}^{(7)}  =  \frac{4\pi^2}{m_b} B_{(i)} \ , \quad
f_{(i)}^{(9)}  =  \frac{16 \pi^2}{m_b} \bigg(A_{(i)}+ \frac{B_{(i)}}2\bigg)\ ,
}
and the Ward Identity is respected if and only if the coefficients $C_{(i)}$ satisfy:
\eq{
C_{(i)} = -\frac{q^2}{m_b} A_{(i)}  - \frac{m_b^2+q^2}{2m_b} B_{(i)} \ .
\label{WI}
}
This condition applies to gauge-invariant combinations and not to single diagrams.
We will detail which are the gauge-invariant combinations below.

\bigskip

We evaluate scalar quantities $A_{(i)},B_{(i)},C_{(i)}$ for all the two-loop diagrams listed in~\Fig{fig:diags},
grouped in different classes $i=\{ a,b,c,d,e\}$, as detailed in the figure.
The results for the functions $A_{(i)},B_{(i)},C_{(i)}$ are given in terms of dimensionless two-loop scalar integrals of the type:
\eq{
j[i;n_{i_1},n_{i_2},n_{i_3},n_{i_4},n_{i_5},n_{i_6},n_{i_7}] = (2\pi)^{-2d}
\int \frac{(m_b^2)^{N_i-4} (\tilde \mu^2)^{2\epsilon}\ d^d\ell \ d^dr}{P_{i_1}^{n_{i_1}} P_{i_2}^{n_{i_2}} P_{i_3}^{n_{i_3}}P_{i_4}^{n_{i_4}} P_{i_5}^{n_{i_5}}P_{i_6}^{n_{i_6}}P_{i_7}^{n_{i_7}}}
}
where the numbers $n_i$ are integers (positive or negative), with $N_i = \sum_{j=1}^7 n_{i_j}$, the objects $P_i$ are propagators (see below),
and the indices $\{i_1,\dots,i_7\}$ depend on the class.
In addition, $d=4-2\epsilon$, and $\tilde \mu^2 \equiv \mu^2 e^{\gamma_E}/4\pi$,
with $\mu$ the $\overline {\rm MS}$ scale.
Our choice of momentum routings fixes the first five propagators in each class, and the other two are chosen to be linear in loop momenta and such that the seven propagators form a linearly-independent set.
The complete set of propagators needed is:
\begin{align}
P_1 &= (\ell+q)^2 - m_c^2 & 
P_5 &= (r+p-q)^2 & 
P_{9}&=\ell\cdot q
\nonumber\\
P_2 &= \ell^2 - m_c^2  & 
P_6 &= r\cdot q   &  
P_{10}&=(r+p-q)^2 - m_b^2
\nonumber\\ 
P_3 &= (\ell+r)^2 - m_c^2  &
P_7 &= \ell \cdot (p-q) &
P_{11}&=  (r+p)^2 - m_b^2
\\
P_4 &= r^2  & 
P_8 &= (r+p)^2   &  
P_{12} &= (\ell+r+q)^2 - m_c^2
\nonumber\\
P_{13} &= r\cdot (p-q)
\nonumber
\end{align}
and the scalar integrals for each class are:

\begin{align}
&j[a;n_2,n_3,n_4,n_5,n_8,n_7,n_{9}]\ , &
j[d;n_1,n_2,n_{12},n_4,n_{11},n_6,n_7]\ ,
\nonumber\\
&j[b;n_2,n_3,n_4,n_{10},n_{11},n_7,n_{9}]\ , &
j[e;n_1,n_2,n_3,n_4,n_{12},n_7,n_{13}]\ ,
\\
&j[c;n_1,n_2,n_3,n_4,n_5,n_6,n_7]\ . &
\nonumber
\end{align}
Once all the two-loop scalar integrals $j[i;\{n_i\}]$ are known, the problem of calculating the invariant functions $f_{(i)}^{(7,9)}$ is solved.
In the following we describe the analytic calculation of the two-loop scalar integrals.

\subsection{IBP reduction and Master integrals}
\label{sec:IBP}

At this point, the result of each diagram is a function of many scalar integrals with many different tuples
$\{n_{i_1},\dots,n_{i_7}\}$ in its class. We can now use integration-by-parts identities (IBPs)
to reduce the set of scalar integrals appearing in each class to a small set of \emph{Master Integrals} (MIs).
For this purpose we use the Mathematica code LiteRed~\cite{Lee:2012cn}. After reduction, the total number of two-loop MIs in each class is $m_i=\{7,9,9,15,5\}$ for $i=\{ a,b,c,d,e\}$, respectively.
These MIs are listed in~\App{app:MIs}, and collectively denoted by $J_{i,k}$,
with $i=\{ a,b,c,d,e\}$, and $k=1\dots m_i$ for each $i$.

With the functions $A_{(i)},B_{(i)},C_{(i)}$ written in terms of MIs one can check the Ward Identity by verifying~\Eq{WI}, which holds analytically and explicitly in terms of the unevaluated MIs.
This does not happen individually for each diagram, but for the following combinations: $a_1$, $a_2+a_3$, $b_1+b_2+b_3$ (only if $Q_s=Q_b$), $c_1+c_2$, $d_1+d_2$, and $e_1+e_2+e_3$, according to the numberings in~\Fig{fig:diags}.

\bigskip

We now perform some simplifying operations on the master integrals.
First, we express the integrands themselves in terms of the invariant variables on which the scalar integrals depend, which we choose to be
\eq{
s \equiv q^2/m_b^2\ ,\quad
z \equiv m_c^2/m_b^2\ .
}
For this purpose we note that there always exist two light-like vectors $k_{1,2}$ ($k_1^2=k_2^2=0$) such that $p=k_1+k_2$ and
$q=k_1 + s \,k_2$. Then $p-q = (1-s) \,k_2$, and the condition $(p-q)^2=0$ is automatically satisfied.
In addition, $k_1\cdot k_2 = m_b^2/2$. Thus, expressing the integrands in terms of $k_{1,2}$ instead of $p,q$ leads to the (dimensionless) scalar integrals $j[i;\{n_i\}]$ as explicit functions of $(s,z)$.

Second, in order to be able to do a rational transformation to a canonical basis of master integrals (as explained below), for each set of diagrams $(i)$ we make a change of variables
$(s ,z) \mapsto (x_i,y_i)$, with $x_i=x_i(s,z)$ and $y_i=y_i(s,z)$ a set of functions that will be specified later. In terms of these new variables the dimensionless MIs are written as
$J_{i,k}(\epsilon,x_i,y_i)$.

\subsection{Differential Equations in canonical form and iterative solution}

For each set of diagrams, we construct the system of differential equations:
\eq{
\partial_x\,J_{i,k}(\epsilon,x,y) =
a_{i,x}^{k\ell}(\epsilon,x,y)\, J_{i,\ell}(\epsilon,x,y)\ ,\quad
\partial_y\,J_{i,k}(\epsilon,x,y) =
a_{i,y}^{k\ell}(\epsilon,x,y)\, J_{i,\ell}(\epsilon,x,y)\ ,
}
where $a_{i,x}$, $a_{i,y}$ are $m_i\times m_i$ matrices
depending on $\epsilon$, $x$ and $y$.
The derivatives of the MIs $J_{i,k}$ are performed by differentiating the integrands, which produce new scalar integrals, and then applying the IBP reduction again on these scalar integrals to express the derivatives $\partial_{x,y}\,J_{i,k}$ themselves in terms of the MIs $J_{i,k}$. One can then read off the matrices $a_{i,x}$ and $a_{i,y}$.

\bigskip

A basis of Master Integrals is said to be ``canonical''~\cite{Henn:2013pwa} if $a_{x,y}(\epsilon,x,y) = \epsilon A_{x,y}(x,y)$, with $A_x(x,y)$ and
$A_y(x,y)$ two $N\times N$ matrices independent of $\epsilon$. Given a canonical basis $\vec M$, the differential equations
have the form:
%
%\eq{
%d \vec M(\epsilon,x,y) = \epsilon\ d \tilde A(x,y)\, \vec M(\epsilon,x,y) 
%}
%
\eq{
\partial_x \vec M(\epsilon,x,y) = \epsilon\  A_x(x,y)\, \vec M(\epsilon,x,y) \quad ; \quad
\partial_y \vec M(\epsilon,x,y) = \epsilon\  A_y(x,y)\, \vec M(\epsilon,x,y) \ .
}
%\eq{
%d \vec M(\epsilon,x,y) = \epsilon\ d \tilde A(x,y)\, \vec M(\epsilon,x,y) 
%}
%
Although not explicitly used in the following, we note that there is a matrix $\tilde A(x,y)$ such that $\partial_x \tilde A(x,y) = A_x(x,y)$ and $\partial_y \tilde A(x,y) = A_y(x,y)$.
%and always exists (so $d \tilde A(x,y) = A_x(x,y) dx + A_y(x,y) dy$).

\bigskip

Once a canonical basis is found, the system of differential equations
can be solved automatically order by order in $\epsilon$.
To keep the notation as simple as possible in this section, we will assume that all the master integrals in the canonical basis are regular in $\epsilon$ (if not, we redefine them by multiplying all of them with the same appropriate power of $\epsilon$).
We then write the $\epsilon$-expansion for the master integrals
\eq{
\vec M(\epsilon,x,y) = \sum_{n=0}^\infty \epsilon^n\, \vec M_n(x,y)
}
and the differential equations read:
\eq{
\partial_{x,y} \vec M_n(x,y) = A_{x,y}(x,y) \vec M_{n-1}(x,y) \ .
\label{eq:diffeqcanonical}
}
We first construct the general solution of the differential equation containing the derivative with respect to $y$.
Using partial fraction decomposition, $A_y$ can be written in the form 
\eq{
A_y(x,y) = \sum_j  \frac{A_y^j}{y-w_j(x)}\ ,
}
where $A_{y}^j$ a set of constant matrices, and the quantities $w_j(x)$ are called the ``$x$-dependent weights" (see~\App{app:weights}).
This differential equations can be solved iteratively due to the structure of (\ref{eq:diffeqcanonical}):
\eqa{
\vec M_0(x,y) & = & \vec C_0(x) \ , 
\nonumber\\[2mm]
\vec M_1(x,y) & = &
\sum_{j_1} \big[A_y^{j_1}\, G(w_{j_1}(x);y)\big] \,\vec C_0(x) + \vec C_1(x) \ ,
\nonumber\\[2mm]
\vec M_2(x,y) & = &
\sum_{j_2,j_1} \big[A_y^{j_2} \, A_y^{j_1} \, G(w_{j_2}(x),w_{j_1}(x);y)\big] \, \vec C_0(x)  +  \sum_{j_2} \big[A_y^{j_2}\, G(w_{j_2}(x);y)\big]  \, \vec C_1(x) + \vec C_2(x) \ ,
\nonumber\\[2mm]
\vec M_3(x,y) & = & \cdots
}
etc., in terms of Generalized Polylogarithms (GPLs)~\cite{Goncharov:1998kja}, defined iteratively as~\cite{Frellesvig:2016ske}
\eq{
G(w_1,\dots,w_n;y)=\int_0^y \frac{dt}{t-w_1} G(w_2,\dots,w_n;t)\ ;\quad
G(;y)=1\ ;\quad
G(\vec 0_n;x) = \frac{\log^n x}{n!}\ ,
}
where $\vec 0_n$ denotes $n$ consecutive zeroes. In each step of the iteration, integration constants (with respect to the $y$ integration) are added, which however depend on the
variable $x$; they are denoted as $\vec C_n(x)$. 

Using the fact that the GPLs in the above equations either tend to zero in the limit $y \to 0$ or to $\frac{\log^n x}{n!}$ (when all $n$ weights are zero), it is straightforward
to derive ordinary differential equations for the $\vec C_n(x)$ quantities, obtaining 
\eq{
\partial_{x} \vec C_n(x) = A_{x}(x,y=0) \, \vec C_{n-1}(x) \ .
\label{eq:diffeqcanonicalx}
}
The matrix $A_x$ evaluated at $y=0$ has, after partial fraction decomposition,
the form
\eq{
A_x(x,y=0) = \sum_k A_x^k \frac{1}{x-w_k}\ ,
}
where $A_{x}^k$ is again a set of constant matrices, and the quantities $w_k$ are now constant weights  (see~\App{app:weights}). The solutions of the differential equations for $\vec C_n(x)$ again are determined iteratively:
\eqa{
\vec C_0(x) & = & \vec C_0 \ , 
\nonumber\\[2mm]
\vec C_1(x) & = &
\sum_{k_1} \big[A_x^{k_1}\, G(w_{k_1};x)\big] \, \vec C_0 + \vec C_1 \ , 
\nonumber\\[2mm]
\vec C_2(x) & = &
\sum_{k_2,k_1} \big[A_x^{k_2} \, A_x^{k_1} \, G(w_{k_2},w_{k_1};x)\big] \, \vec C_0  +  \sum_{k_2} \big[A_x^{k_2}\, G(w_{k_2};x)\big] C_1 + \vec C_2  \ ,
\nonumber\\[2mm]
\vec C_3(x) & = & \cdots
}
where $\vec C_n$ on the right-hand side are constants with respect to both variables.

Thus, the problem of calculating the MIs is reduced to find a canonical basis and to fix the integration constants, which is a much more tractable challenge.
In order to find a canonical basis for each set $J_{i,k}$ of MIs, we use the mathematica program {\it CANONICA}~\cite{Meyer:2017joq}. This code is able to look for transformations that involve \emph{rational} functions of the arguments. For this reason, the right set of variables $(x_i,y_i)$ must be found for each case before using this program. Starting from our original variables $s=q^2/m_b^2$ and $z=m_c^2/m_b^2$, we define, for each diagram set $i$,
the variables $x_i$ and $y_i$:
\eqa{
&&
x_a=x_c=x_e=\frac{1}{\sqrt{1-4z}}\ ,\quad
x_b=x_d=\sqrt{4z}-\sqrt{4z-1}\ ,
\nonumber\\[2mm]
&&
y_a=\frac{1}{\sqrt{1-\frac{4z}{1-s}}} \ , \quad 
y_b=\frac{1}{\sqrt{1-\frac{4}{s}}} \ , \quad 
y_c=y_d=y_e=\frac{1}{\sqrt{1-\frac{4z}{s}}} \ .
\label{eq:xiyi}
}
In terms of these variables and with the help of {\it CANONICA}, we are able to find linear transformations
\eq{
M_{i,k} = (T_{i}^{-1})^{k\ell}(\epsilon,x_i,y_i)\, J_{i,\ell}
\label{eq:T}
}
such that the MIs $M_{i,k}$ constitute a canonical basis for each set $i=\{a,c,d,e\}$. For set $b$, the situation is somewhat more complicated: There is a linear transformation involving rational functions of the arguments $x_b$ and $y_b$ for the MIs $J_{b,1-6}$ and this six-dimensional block can be treated in a straightforward way, but the complete nine-dimensional problem contains complicated square roots of these variables in the transformation matrix to the
canonical basis and in the matrices $A_x$ and $A_y$ which define the differential equations in this basis. Similar as after Eq.~(4.46) of~\Ref{Bell:2014zya}, we introduced the variables $t_b$ and $v_b$ to rationalize these roots:
\eqa{
t_b&=&
\frac{-4 x_b^2 + 4 x_b^2 y_b + 2\sqrt2 x_b^2  (1 + y_b)
\sqrt{
\frac{2 x_b^4 - x_b^2 y_b + 2 x_b^4 y_b - x_b^6 y_b + x_b^2 y_b^2 + 4 x_b^4 y_b^2 + x_b^6 y_b^2}{x_b^4 (1 + y_b)^2}
}}{-1 + 6 x_b^2 - x_b^4 + y_b + 2 x_b^2 y_b + x_b^4 y_b}\ ,
\nonumber\\[4mm]
v_b&=& \frac{-4 x_b^2 - 4 x_b^2 y_b + 4\sqrt2 x_b^2  (1 -y_b) \sqrt{
\frac{2 x_b^4 + x_b^2 y_b - 2 x_b^4 y_b + x_b^6 y_b + x_b^2 y_b^2 + 4 x_b^4 y_b^2 + x_b^6 y_b^2}{x_b^4 (1 - y_b)^2}
}}{1 - 6 x_b^2 + x_b^4 + y_b + 2 x_b^2 y_b + x_b^4 y_b}\ .
\label{eq:tbvb}
}
For this reason the results for the MIs $J_{b,7}$, $J_{b,8}$, and $J_{b,9}$ involve GPLs with arguments $t_b$ and/or $v_b$.

We stress that the chosen variables $x_i$ have the properties that they tend to zero when $z$ goes to infinity. Similarly, the variables $y_i$ (as well as $t_b$ and $v_b$) go to zero for $s \to 0$ (when $i=b,c,d,e$) and for $s \to 1$ (when $i=a$), independently of the value of $z$. In these limits, the functions $G(...;x_i)$, $G(...;y_i)$, $G(...;t_b)$ and $G(...;v_b)$ can be expanded in a straightforward way for the small values of $x_i$, $y_i$, $t_b$ and $v_b$, respectively. 
This turns out to be very useful when fixing the integration constants in the following section, because we will heavily make use of the asymptotic properties of the originals integrals $J_{i,k}$
in the limit where $x_i$ and/or $y_i,t_b,v_b$ go to zero.

\subsection{Fixing integration constants and analytic continuation}
\label{sec:BoundaryConditions}

Once the canonical basis is found and the general solution of the differential equations in this basis is constructed, we have to fix the integration constants. 
To this end we transform in a first step the MIs back to the original basis by making use of the transformation matrices~$T_i$ (i.e.~\Eq{eq:T}).
The constants are then determined by either computing the MIs $J_{i,k}$ in the various classes $i$ at a particular kinematical point for which the calculation is simple, or by using
asymptotic properties in the limit $z \to \infty$. These properties follow in a straightforward way from the heavy mass expansion (HME) of a given integral \cite{Smirnov:1994tg}.

We explain this in some detail for the nine MIs in class $c$: it turns out that only the integral~$J_{c,1}$, which is simply a product of two one-loop tadpole integrals, has to be calculated
explicitly. In the limit for large $m_c$ ($m_c \gg m_b$) the other eight integrals can be naively Taylor expanded in the external momenta and in $m_b$. Note that in the present situation the only subdiagrams in the sense of the HME are just the full diagrams (i.e. the full MIs) and therefore the naive Taylor expansion is justified. The leading power $n$ in the $m_c$-expansion of a given integral $J$ is then identical to the
mass dimension of the integral, where the mass dimension is an even integer; the structure of $J$ is
\eqa{
J=K \, m_c^n\, P(q^2/m_c^2,m_b^2/m_c^2) \, ,
}
where $K$ is a constant prefactor and $P$ is a polynomial of the indicated arguments.

The GPLs in the general solution for the MIs (from the differential equations) can be easily expanded for large $z$ and small $s$ in class $c$. Very often, the expanded solution for a given integral contains higher powers in $m_c$ than that determined from the HME argumentation. The requirement that these terms are absent allows to determine some of the integration constants.
From the HME structure it is also clear that only even powers of $m_c$ can be present; this fact fixes the remaining integration constants. It is worth emphasizing that all constants can be fixed by the explicit knowledge $J_{c,1}$ in class $c$ and the structure of the powers in $m_c$.
{\it The explicit HME evaluation of the MIs is not even necessary.}

For classes $\{b,d,e\}$ the fixing of the integration constants is done in the same way as in class~$c$: only a small number of simple one-loop integrals have to be calculated explicitly;
again the GPLs in the results for the MIs (from the differential equations) can be easily expanded for large $z$ and small $s$ and all constants can be fixed.
A few examples on the fixing of integration constants in classes $c$ and $e$ are given in~\App{app:ExamplesBCs}.

We now turn to class $a$.
Due to the variable $y_a=1/\sqrt{1-4z/(1-s)}$, we need to use the behavior of the solutions of the MIs near $s = 1$
(not at $s = 0$ as in the other classes) and again for $z \to \infty$. 
Apart from heavy mass expansion arguments (which are the same as in the other classes), we need to calculate directly the three integrals $J_{a,1}$, $J_{a,4}$ and $J_{a,5}$ (which all factorize into two one-loop integrals), in order to fix the integration constants.
Among them, only $J_{a,4}$ depends on $s$. The explicit result reads
\begin{equation}
J_{a,4}=
 \frac{e^{2\epsilon \gamma_E}}{(4 \pi)^{4}}\frac
{\Gamma(\epsilon-1) \Gamma(\epsilon) \Gamma(1-\epsilon)^2}
{\Gamma(2-2 \epsilon)} (\mu/m_b)^{4 \epsilon} z^{1-\epsilon} (-s)^{-\epsilon}\ .
\label{eq:Ja4}
\end{equation}
When expanding this result in $\epsilon$, $\log(-s)$ appears where $s$ is understood to have a small positive imaginary part in order to properly represent
the original Feynman integral. The result (\ref{eq:Ja4}) is therefore just the analytic continuation of the Feynman integral onto the complex plane cut along the positive real $s$-axis, having a discontinuity on this axis. 
However, when expanding the GPLs in the solution of the differential equations for $J_{a,4}$ around $s=1$, we find a regular behavior, which is due to the fact that the solution in terms of GPLs with argument $y_a$ represents a different analytic continuation. In order to obtain an analytic continuation with the branch cut along the positive real $s$-axis (see~Sections~\ref{sec:AnalStr} and~\ref{sec:analytic}) we need to consider the differential equations for the upper and the lower $s$-half planes separately. In particular, we have to {\it fix the integration constants for the two pieces separately.}
In this way, the branch cuts in all classes appear along the positive real axis, starting at $s=\{0,4z,4\}$, depending on the class. These branch cuts will be analyzed in detail in~\Sec{sec:analytic}.

Our final results for all MIs in the Feynman region have been checked numerically using Sector Decomposition as implemented in SecDec~\cite{Carter:2010hi,Borowka:2015mxa}.

\subsection{Counterterm contributions}
\label{subsec:countertermso2}

For the renormalization we will follow closely~\Ref{Asatryan:2001zw}, and therefore we prefer to stick to the notation of that paper within this section:
\eq{
O_{1,2} \equiv\cO_{1,2}\ ;
\qquad \widetilde O_{7,9} \equiv \cO_{7,9} \ ;
\qquad O_{7,9} \equiv \frac{4\pi}{\alpha_s} \cO_{7,9}\ .
}
In~\Ref{Asatryan:2001zw} the final results were written as linear combinations of the tree-level matrix elements of $\widetilde O_7$ and $\widetilde O_9$.
In this section we generalize the formulas of~\Ref{Asatryan:2001zw} to hold for arbitrary values
of the squared momentum transfer $q^2$ and write the results in terms of $\langle {\cal O}_7 \rangle_{\rm{tree}}$ and $\langle {\cal O}_9 \rangle_{\rm{tree}}$, as in~\Eq{eq:<CiOi>}.

Up to this point we have calculated the bare two-loop contributions to $\Delta C_{7,9}$ from the diagrams in~\Fig{fig:diags}. As the operators $\cO_{1,2}$ mix under renormalization, there are additional contributions at order $\cO(\alpha_s)$ proportional to $C_{1,2}$. These counterterm contributions arise from the matrix elements of the operators
\eq{
\sum_{j=1}^{12}
\delta Z_{ij} O_j \ ,\quad i=1,2\ . 
}
The set of operators $O_1$--$O_{10}$ is given in Eq.~(2) of~\Ref{Asatryan:2001zw},
while $O_{11}$ and $O_{12}$ are evanescent, that is, they vanish in $d=4$ dimensions.
Although there is certain freedom in the choice of the evanescent operators (e.g. one may add terms of order $\epsilon$), it is convenient to use the same definitions as in~\Ref{Bobeth:2000mk} in order to combine our matrix elements with the  Wilson coefficients calculated there:
\eqa{
O_{11} &=&
\big( \bar{s}_L \gamma_\mu \gamma_\nu \gamma_\sigma T^a c_L \big)
\big( \overline{c}_L \gamma^\mu \gamma^\nu \gamma^\sigma T^a b_L \big)
- 16 \, O_1 \, ,
\\
O_{12} &=&
\big( \bar{s}_L \gamma_\mu \gamma_\nu \gamma_\sigma c_L \big)
\big( \overline{c}_L \gamma^\mu \gamma^\nu \gamma^\sigma b_L \big)
- 16 \, O_2 \, .
}
The renormalization constants $\delta Z_{ij}$ are written as
\eq{
\delta Z_{ij} = \frac{\alpha_s}{4 \pi} \left( a_{ij}^{01} +
\frac{1}{\epsilon} a_{ij}^{11}\right) + \frac{\alpha_s^2}{(4 \pi)^2}
\left( a_{ij}^{02} + \frac{1}{\epsilon} a_{ij}^{12} + \frac{1}{\epsilon^2} a_{ij}^{22}\right)
+ \cO(\alpha_s^3)\ ,
}
with the relevant coefficients~\cite{Bobeth:2000mk,Asatryan:2001zw}
\eq{
    \hat{a}^{11}=
    \left(
    \begin{array}{cccccccccccc}
        -2& \frac{4}{3}&0& -\frac{1}{9} &0&0 & 0 &0& -
\frac{16}{27}&0&\frac{5}{12}& \frac{2}{9} \\[3mm]
         6& 0&0 & \frac{2}{3} &0&0&0 &0& -\frac{4}{9}& 0& 1& 0
    \end{array}
    \right)\,,\ 
    \begin{array}{lll}
        a^{12}_{17} = -\frac{58}{243}\,,\  &
        a^{12}_{19} = -\frac{64}{729}\,,\  &
        a^{22}_{19} = \frac{1168}{243}\,,\ \\[3mm]
        a^{12}_{27} = \frac{116}{81}\,,\  &
        a^{12}_{29} = \frac{776}{243}\,,\  &
        a^{22}_{29} = \frac{148}{81}\,.
    \end{array}
}
The counterterm contributions to the functions $F_i^{(7,9)}$ due to the mixing of $O_{1,2}$ into four-quark operators are denoted by $F_{i \to \rm{4\,quark}}^{\rm{ct}(7,9)}$, and are related to the one-loop matrix elements of four-quark operators by
\eq{
\label{fi4quark}
\sum_{j} \left( \frac{\alpha_s}{4\pi}\right) \, \frac{1}{\epsilon} \,
a_{ij}^{11} \langle s \ell^+ \ell^-|O_j|b \rangle_{\text{1-loop}}
= - \left( \frac{\alpha_s}{4\pi}\right) \, \left[
 F_{i \to \rm{4 quark}}^{\rm{ct}(7)} \langle {\cal O}_7\rangle_{\text{tree}} +
 F_{i \to \rm{4 quark}}^{\rm{ct}(9)} \langle {\cal O}_9\rangle_{\text{tree}}
\right] \, ,
}
where $j$ runs over the set of four-quark operators. Since many entries of $\hat{a}^{11}$ are zero, only the one-loop matrix elements of $O_1$, $O_2$,  $O_4$, $O_{11}$ and $O_{12}$ are needed. These matrix elements are needed to order $\epsilon^1$. Compared to~\Ref{Asatryan:2001zw}, we worked out the exact results, expressed in terms of GPLs. 

The counterterm contributions from the mixing of $O_i$ ($i=1,2$) onto $O_9$ are of two types:
The first type corresponds to the one-loop mixing $O_i \to O_9$, followed by taking the one-loop matrix element of $O_9$. This contributes to the renormalization of the diagram on the right hand side in \Fig{fig:diagsIR} and does not contribute to the functions $F_i^{(j)}$. 
The second type is due to (a) the two loop mixing of $O_i \to O_9$ and (b) the one-loop mixing combined with the one-loop renormalization of the $\alpha_s$ factor in the definition of the operator $O_9$.
The corresponding contributions to the form factors are denoted by $F_{i  \to 9}^{\rm{ct}(7,9)}$, and given by~\cite{Asatryan:2001zw}
\eq{
F_{i  \to 9}^{\rm{ct}(7)} = 0 \, ;
\quad
F_{i  \to 9}^{\rm{ct}(9)} = -\left(\frac{a_{i9}^{22}}{\epsilon^2} + \frac{a_{i9}^{12}}{\epsilon} \right) - \frac{a_{i9}^{11} \, \beta_0}{\epsilon^2} \ . 
}
for which the strong coupling renormalization constant $Z_{g_s}$ is needed:
\eq{
Z_{g_s} = 1 - \frac{\alpha_s}{4\pi} \, \frac{\beta_0}2 \, \frac1{\epsilon} \ ;
\quad
\beta_0 = 11 - \frac23 n_f \ ;
\quad
n_f = 5 \ .
}
The contributions generated by the two-loop mixing of $O_1$ and $O_2$ into $O_7$ are given by
\eq{
F_{i  \to 7}^{\rm{ct}(7)} = - \frac{a_{i7}^{12}}{\epsilon} \ ;
\quad 
F_{i  \to 7}^{\rm{ct}(9)} = 0 \ .
}

In addition to the contributions from operator mixing, there is a contribution from the renormalization of the charm quark mass. This is taken into account by replacing $m_c$ with $Z_{m_c} \cdot m_c$ in the one loop contributions given in~\Eq{eq:f9LO}.
Note that in this paper we are using the pole mass definition of $m_c$, characterized by the renormalization constant
\eq{
Z_m = 1 -\frac{\alpha_s}{4\pi}\, C_F \left( \frac{3}{\epsilon} + 6 \log
\frac{\mu}{m_c} + 4 \right) .
}

We have checked that the sum of the divergent parts of all these counterterm contributions is identically opposite to that of the unrenormalized matrix elements, thus proving the cancellation of ultraviolet divergences. 

On the other hand, the finite part of the counterterm contributions, which we denote by $F_i^{\text{ct}(j)}$  ($i=1,2$; $j=7,9$), contribute to the renormalized NLO functions~$F_i^{(j)}$.
Besides working out the exact results for the counterterm contributions $F_i^{\text{ct}(j)}$ in terms of GPLs, we have also separated the different contributions proportional to the different charge factors $Q_{s,c,b}$, since they renormalize the different contributions to $F_i^{(j)}$ with different analytic structure. It turns out that the only contributions proportional to $Q_{s,b}$ to $F_i^{\text{ct}(j)}$ come from the mixing $O_i \to O_4$, specifically from the one-loop matrix element of $O_4$ with an $s$- or $b$-quark in the loop, and thus these contributions are easy to isolate. In the end, our results for the counterterm contributions are given by the sum of three pieces:
\eq{
F_i^{\text{ct}(j)} = F_{i,Q_s}^{\text{ct}(j)} + F_{i,Q_c}^{\text{ct}(j)} + F_{i,Q_b}^{\text{ct}(j)}\ .
}
with $i=\{1,2\}$; $j=\{7,9\}$.
All these functions are given separately in electronic form in an ancillary file (c.f.~\App{app:F279}).

\subsection{Results for renormalized matching coefficients at NLO}
\label{sec:NLOresults}

Collecting all the pieces, the final results for the matching coefficients $\Delta C_{7,9}(q^2)$ in~\Eq{eq:HOPE} at NLO are given by
\eqa{
\Delta C_7(q^2) &=& - \frac{\alpha_s}{4\pi} \Big[ C_1 F_1^{(7)}(q^2) +C_2 F_2^{(7)}(q^2)\Big] + \cO(\alpha_s^2)\ ,
\label{eq:MatchingC7}
\\[3mm]
\Delta C_9(q^2) &=& 
f^{(9)}_{\rm LO}(q^2) - \frac{\alpha_s}{4\pi} \Big[
C_1 F_1^{(9)}(q^2) +C_2 F_2^{(9)}(q^2)
\Big]+ \cO(\alpha_s^2)
 \ ,
\label{eq:MatchingC9}
}
where $f^{(9)}_{\rm LO}(q^2)$ is given in~\Eq{eq:f9LO} and the renormalized NLO functions $F_{1,2}^{(7,9)}(q^2)$ are the sum of the contributions from the two-loop diagrams $a$ through $e$ and the counterterm contributions:
\eq{
F_i^{(j)} = F_{i(a)}^{(j)} + F_{i(b)}^{(j)}
+ F_{i(c)}^{(j)} + F_{i(d)}^{(j)} + F_{i(e)}^{(j)}
+ F_{i}^{{\rm ct}(j)}\ ,
}
with $i=\{1,2\}$; $j=\{7,9\}$.
The functions $F_{1(\rm diag)}^{(j)}$ are related to $F_{2(\rm diag)}^{(j)}$ by a simple color factor, depending on the diagram:
\eq{
F_{1(a,b,c,d)}^{j} = -\frac1{2N_c}  F_{2(a,b,c,d)}^{j}\ , \quad
F_{1(e)}^{j} = C_F\,  F_{2(e)}^{j}\ .
}

The complete analytic results for the functions $F_{i(k)}^{(j)}(q^2)$ --with $i=\{1,2\}$, $j=\{7,9\}$ and $k=\{a,b,c,d,e\}$--, $F_{i}^{{\rm ct}(j)}(q^2)$, and the full $F_{i}^{(j)}(q^2)$ are given in electronic form in an ancillary \texttt{Mathematica} package attached to the arXiv submission of this paper. See~\App{app:F279} for details. The attached program is the same that we have used for all the numerics in the following sections.

The coefficients $\Delta C_{7,9}(q^2)$ can also be split in the two different contributions $\Delta C^{(c)}_{7,9}(q^2)$ and $\Delta C^{(sb)}_{7,9}(q^2)$ proportional to the charge factors $Q_c$ and $Q_{s,b}$ respectively, and contributing to the functions $\H^{\rm OPE}_{\lambda,c}(q^2)$ and $\H^{\rm OPE}_{\lambda,sb}(q^2)$ discussed in~\Sec{sec:AnalStr}. For this separation we refer to~\Sec{sec:analytic} below.

\section{Numerical evaluation of NLO corrections}
\label{sec:numerics}
\setcounter{equation}{0}

\subsection{Numerical evaluation of GPLs}
\label{sec:NumGPLs}

For the fast numerical evaluation of the GPLs we use the C\texttt{++} ginac package~\cite{ginac} interfaced with \texttt{Mathematica}.
In particular, we use the ginac multiple polylogarithm \texttt{G}, to evaluate the GPLs with unit argument and the last weight non-zero:
\eq{
G(w_1,\dots,w_n;1)\ , \quad \text{with}\ w_n\ne 0\ .
}
When $w_n\ne 0$, the GPL with arbitrary (non-zero) argument is obtained from the identity
\eq{
G(w_1,\dots,w_n;x) = G\Big(\frac{w_1}{x},\dots,\frac{w_n}{x};1\Big)\ , \quad \text{if}\ w_n,x\ne 0\ ,
\label{eq:4.2}
}
while the GPL with zero argument is zero. This part is implemented by the \texttt{Mathematica} interface. In order to evaluate the cases with $w_n=0$ we need to eliminate all the ``trailing zeroes" in the GPLs, which refer to any string of consecutive zeroes at the end of the weight list, e.g.,
$G(1,-2 i,0,0\,;3+ i)$ has two trailing zeroes. Reexpressing the GPLs in terms of new GPLs without trailing zeroes is also done by the \texttt{Mathematica} interface, recursively in the number of trailing zeroes, by means of the following formula:
\eqa{
&&\hspace{-10mm}
G(w_1,\dots,w_n,\underbrace{0\dots 0}_{m};x) = \frac1{m} \bigg[
\log{x}\ G(w_1,\dots,w_n,\underbrace{0\dots 0}_{m-1};x)
-G(0,w_1,\dots,w_n,\underbrace{0\dots 0}_{m-1};x)
\nonumber\\[2mm]
&&\hspace{5mm}
-G(w_1,0,w_2\dots,w_n,\underbrace{0\dots 0}_{m-1};x)
- \cdots
-G(w_1,\dots,w_{n-1},0,w_n,\underbrace{0\dots 0}_{m-1};x)
\bigg]\ .
\label{eq:4.3}
}
This provides a complete algorithm for the evaluation of any GPL.
For convenience, we provide our C\texttt{++}/\texttt{Mathematica}
bundle (with front-end package \texttt{GPL.m}) as an ancillary file supplementing this paper (see \App{app:GPLs} for details).
All our numerical results have also been reproduced using \texttt{Maple},
which includes a built-in function for GPLs. However, the evaluation within \texttt{Maple} is significantly slower that the one provided by \texttt{GPL.m}. 

In order to properly evaluate our expressions, we consider separately the GPLs with arguments $x_i$ or $y_i$. For GPLs with argument $x_i$, we numerically evaluate $x_i$ by adding a small negative imaginary part to $z$, typically of order $10^{-12}$. For GPLs with argument $y_i$, on the contrary, we evaluate the $x_i$ dependent weights in the limit in which the small imaginary part on $z$ tends to zero; the arguments $y_i$ are calculated by taking $z$ real from the beginning and by adding a small positive/negative imaginary part (typically of order $10^{-8}$) to $s$, when $s$ lies on the real axis.

\subsection{Numerical evaluation of NLO corrections and tests}

Once the numerical evaluation of the GLPs has been addressed, the numerical evaluation of the NLO functions $F_i^{j}(s,z)$ is relatively simple. We use the \texttt{Mathematica} package \texttt{FFNLO.m}, which is attached to the arXiv submission of this paper (see~\App{app:F279} for details). This program makes a prior list of all the GPLs appearing in the functions to be evaluated, evaluates them only once using \texttt{GPL.m}, and then substitutes the values in the functions. In addition, it takes into account the sign of ${\rm Im}(s)$ correctly, as the functions $F_{i(a)}^{j}$  have a different form in the upper or lower complex-s plane due to the double fixing of boundary conditions (i.e.~\Sec{sec:BoundaryConditions}). The prescription for $z$ is fixed as described above.

\begin{figure}
\includegraphics[width=\textwidth]{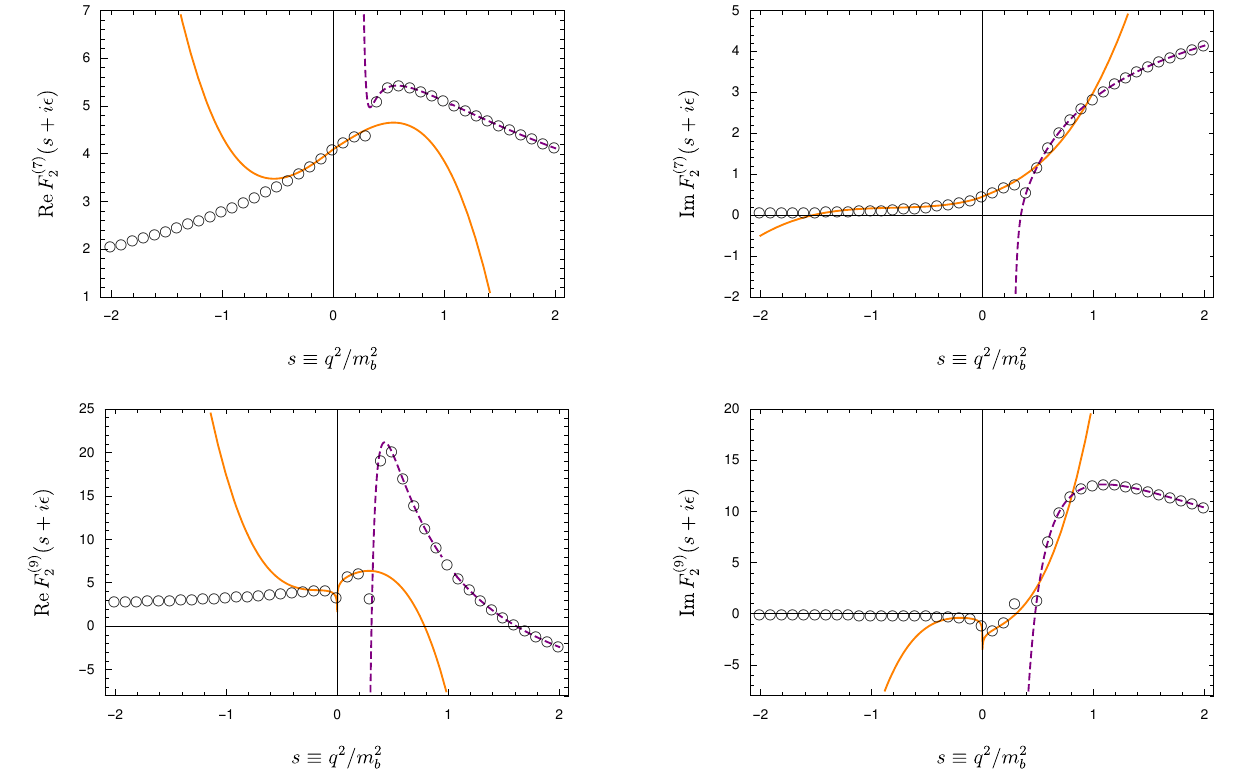}
\caption{Comparison of our exact results (black circles), with the expanded results of~\Ref{Asatryan:2001zw} at low-$q^2$ (solid orange line) and the ones of~\Ref{Greub:2008cy} at high-$q^2$ (dashed purple line).
Note that we have plotted the results of~Refs.~\cite{Asatryan:2001zw,Greub:2008cy} beyond their region of applicability. In these plots we have set $z=(0.29)^2$ and $\epsilon=10^{-8}$.}
\label{fig:comparison}
\end{figure}

We have tested the results against those in~Refs.~\cite{Asatryan:2001zw,Greub:2008cy}, finding very good numerical agreement with Tables~1 and~2 in both papers.
As already mentioned, the results of~Refs.~\cite{Asatryan:2001zw,Greub:2008cy} apply specifically to the low-$q^2$ and high-$q^2$ regions respectively. In \Fig{fig:comparison} we have plotted these results within and beyond their respective regions of applicability and compared them with the analytic results obtained in this paper. We find an excellent agreement within the appropriate regions.
Deviations with respect to the low-$q^2$ results occur starting around $s\lesssim-0.4$. Thus, for the calculation of the OPE matching coefficients in this region it may be advisable to use the results given in the present paper.

\subsection{Selected results at different values of $s$ and $z$}

The results for the NLO functions $F_{1,2}^{(7,9)}(q^2)$ are intended to be used to calculate the function $\H_\mu$ in the OPE region, by means of~Eqs.~(\ref{eq:HOPE}), (\ref{eq:MatchingC7}) and (\ref{eq:MatchingC9}). For the determination of exclusive $b\to s\ell\ell$ amplitudes at large hadronic recoil, this OPE region corresponds to the region of negative $q^2$~\cite{Khodjamirian:2012rm,Bobeth:2017vxj}.
For reference we collect, in~\Tab{tab:valuesnegativeq2}, numerical values for the NLO functions at the points $s=\{-0.6,-0.5,-0.4,-0.3,-0.2,-0.1\}$ for three values of the charm mass, $z=\{(0.25)^2,(0.29)^2,(0.33)^2\}$. As the $m_c$ dependence for these values of $s$ is mild, a quadratic interpolation of the values at these three points will represent this dependence accurately enough.

\begin{table}
\centering
\setlength{\tabcolsep}{10pt}
\begin{tabular}{@{}clll@{}}
\toprule[0.7mm]
$s=q^2/m_b^2$  & \hspace{1cm} $z = (0.25)^2$ &  \hspace{1cm} $z = (0.29)^2$  & \hspace{1cm} $z = (0.33)^2$  \\
\midrule[0.7mm]
\multirow{4}{*}{$-0.6$} & $F_1^{(7)} = -0.597 - 0.043\,i$ & $F_1^{(7)} = -0.534-0.028\,i$ & $F_1^{(7)} = -0.472-0.017\,i$ \\
                        & $F_1^{(9)} = -2.962+0.044\,i$ & $F_1^{(9)} = -3.642+0.035\,i$ & $F_1^{(9)} = -4.214+0.024\,i$ \\
                        & $F_2^{(7)} = +3.580+0.257\,i$ & $F_2^{(7)} = +3.206+0.168\,i$ & $F_2^{(7)} = +2.831+0.100\,i$ \\
                        & $F_2^{(9)} = +4.940-0.265\,i$ & $F_2^{(9)} = +3.654-0.207\,i$ & $F_2^{(9)} = +2.511-0.144\,i$ \\
\midrule
\multirow{4}{*}{$-0.5$} & $F_1^{(7)} = -0.620-0.049\,i$ & $F_1^{(7)} = -0.555-0.032\,i$ & $F_1^{(7)} = -0.489-0.019\,i$ \\
                        & $F_1^{(9)} = -3.714+0.047\,i$ & $F_1^{(9)} = -4.364+0.038\,i$ & $F_1^{(9)} = -4.895+0.027\,i$ \\
                        & $F_2^{(7)} = +3.721+0.293\,i$ & $F_2^{(7)} = +3.327+0.192\,i$ & $F_2^{(7)} = +2.935+0.114\,i$ \\
                        & $F_2^{(9)} = +5.180-0.284\,i$ & $F_2^{(9)} = +3.768-0.228\,i$ & $F_2^{(9)} = +2.531-0.162\,i$ \\
\midrule
\multirow{4}{*}{$-0.4$} & $F_1^{(7)} = -0.645-0.056\,i$ & $F_1^{(7)} = -0.576-0.037\,i$ & $F_1^{(7)} = -0.508-0.022\,i$ \\
                        & $F_1^{(9)} = -4.626+0.051\,i$ & $F_1^{(9)} = -5.221+0.043\,i$ & $F_1^{(9)} = -5.688+0.031\,i$ \\
                        & $F_2^{(7)} = +3.872+0.337\,i$ & $F_2^{(7)} = +3.458+0.220\,i$ & $F_2^{(7)} = +3.046+0.131\,i$ \\
                        & $F_2^{(9)} = +5.452-0.306\,i$ & $F_2^{(9)} = +3.887-0.255\,i$ & $F_2^{(9)} = +2.542-0.186\,i$ \\
\midrule
\multirow{4}{*}{$-0.3$} & $F_1^{(7)} = -0.673-0.065\,i$ & $F_1^{(7)} = -0.600-0.043\,i$ & $F_1^{(7)} = -0.528-0.025\,i$ \\
                        & $F_1^{(9)} = -5.763+0.055\,i$ & $F_1^{(9)} = -6.261+0.049\,i$ & $F_1^{(9)} = -6.626+0.036\,i$ \\
                        & $F_2^{(7)} = +4.036+0.392\,i$ & $F_2^{(7)} = +3.599+0.256\,i$ & $F_2^{(7)} = +3.165+0.152\,i$ \\
                        & $F_2^{(9)} = +5.755-0.332\,i$ & $F_2^{(9)} = +4.004-0.292\,i$ & $F_2^{(9)} = +2.531-0.218\,i$ \\
\midrule
\multirow{4}{*}{$-0.2$} & $F_1^{(7)} = -0.702-0.077\,i$ & $F_1^{(7)} = -0.625-0.050\,i$ & $F_1^{(7)} = -0.549-0.030\,i$ \\
                        & $F_1^{(9)} = -7.233+0.062\,i$ & $F_1^{(9)} = -7.556+0.058\,i$ & $F_1^{(9)} = -7.758+0.045\,i$ \\
                        & $F_2^{(7)} = +4.213+0.462\,i$ & $F_2^{(7)} = +3.750+0.302\,i$ & $F_2^{(7)} = +3.293+0.179\,i$ \\
                        & $F_2^{(9)} = +6.079-0.370\,i$ & $F_2^{(9)} = +4.094-0.348\,i$ & $F_2^{(9)} = +2.470-0.269\,i$ \\
\midrule
\multirow{4}{*}{$-0.1$} & $F_1^{(7)} = -0.734-0.092\,i$ & $F_1^{(7)} = -0.652-0.060\,i$ & $F_1^{(7)} = -0.572-0.036\,i$ \\
                        & $F_1^{(9)} = -9.235+0.078\,i$ & $F_1^{(9)} = -9.226+0.078\,i$ & $F_1^{(9)} = -9.154+0.062\,i$ \\
                        & $F_2^{(7)} = +4.404+0.554\,i$ & $F_2^{(7)} = +3.915+0.362\,i$ & $F_2^{(7)} = +3.432+0.215\,i$ \\
                        & $F_2^{(9)} = +6.353-0.465\,i$ & $F_2^{(9)} = +4.072-0.470\,i$ & $F_2^{(9)} = +2.270-0.373\,i$ \\
\bottomrule[0.7mm]
\end{tabular}
\caption{\it Values for the functions $F_{1,2}^{(7,9)}(q^2)$ at negative $q^2$, for three choices of $z=m_c^2/m_b^2$. The renormalization scale has been fixed to $\mu = m_b$. These numbers do not depend on whether one includes an infinitesimal positive or negative imaginary part for $s$.}
\label{tab:valuesnegativeq2}
\end{table}

\section{Study of the analytic structure at NLO}
\label{sec:analytic}

\setcounter{equation}{0}

\subsection{Singularities of the NLO functions}

The matching coefficients $\Delta C_{7,9}(q^2)$ will mimic the analytic structure of the non-local form factors $\H_\lambda(q^2)$ discussed in~\Sec{sec:AnalStr}. In this case the analytic singularities are due to on-shell intermediate partonic states in the $b\to s \ell\ell$ amplitude, producing branch cut discontinuities in both variables $q^2$ and $(q+k)^2$. This structure can be observed explicitly in the analytic results for $\Delta C_{7,9}(q^2)$ calculated here, where the contribution from each diagram to each singularity can be checked. 

The expected singularity structure is the following.
First, the analytic structure of each of the diagrams as a function of complex $s \equiv q^2/m_b^2$ can be chosen to have a branch cut on the positive real line above some specified (perturbative) threshold: $s>s_{\rm th}$, where the threshold depends on the diagram.
In addition, some diagrams are real on the real line below the threshold, while some are complex-valued.
This is due to the fact that some of the diagrams (the ones that are complex) contain on-shell cuts in the variable $p_b^2 \equiv (q+k)^2$, which we fix to $p_b^2= m_b^2$ from the start.
According to their (expected) analytic structure, the set of diagrams can be classified in four groups:

\begin{enumerate}

\item Diagram $b_2$: Branch cut for $s>4$, real for $s<4$.

\item Diagrams $d$ and $e$: Branch cut for $s>4z$, real for $s<4z$.

\item Diagrams $c$: Branch cut for $s>4z$, complex for $s<4z$.

\item Diagram $a_2$: Branch cut for $s>4m_s^2/m_b^2 \simeq 0$, complex for $s<0$.

\end{enumerate}
The rest of the diagrams, $a_{1,3}$ and $b_{1,3}$ do not have branch cuts in the variable $s$ because the
photon couples to the external legs of the diagram.
Note also that the specific threshold ($4m_s^2/m_b^2$, $4m_c^2/m_b^2$ or $4 m_b^2/m_b^2$) can be determined from the charge coupling
(whether the diagram is proportional to $Q_s$, $Q_c$ or $Q_b$). This relates to the discussion in~\Sec{sec:AnalStr}, and applies also to the counterterm contributions.

\bigskip

From the explicit results obtained here for the contribution to $\Delta C_{7,9}$ from each group of diagrams and counterterms, we can check this analytic structure. This is done in two steps:

\begin{enumerate}

\item Checking explicitly that the discontinuity lies where it is expected, and that the values of each contribution below threshold is real or complex as predicted.

\item Checking appropriate dispersion relations, thus supporting the absence of further singularities besides the expected branch cuts. This is done by checking, for each diagram class, the following equation:
\eq{
F_i^{(j)}(s_1) - F_i^{(j)}(s_0) =
\frac{s_1-s_0}{2\pi i} \int_{s_{th}}^\infty dt\  \frac{F_i^{(j)}(t+i0)-F_i^{(j)}(t-i0)}{(t-s_1)(t-s_0)}\ ,
\label{eq:DR}
}
for any two points $\{s_0,s_1\}$ in the complex plane. Any additional singularities will (generically) produce extra contributions beyond the integral in the r.h.s., and thus the fact that this dispersion relation holds is consistent with the absence of additional singularities anywhere on the complex plane, away from the real interval $[s_{\rm th},\infty)$ . 

\end{enumerate}

\begin{figure}
\includegraphics[width=\textwidth]{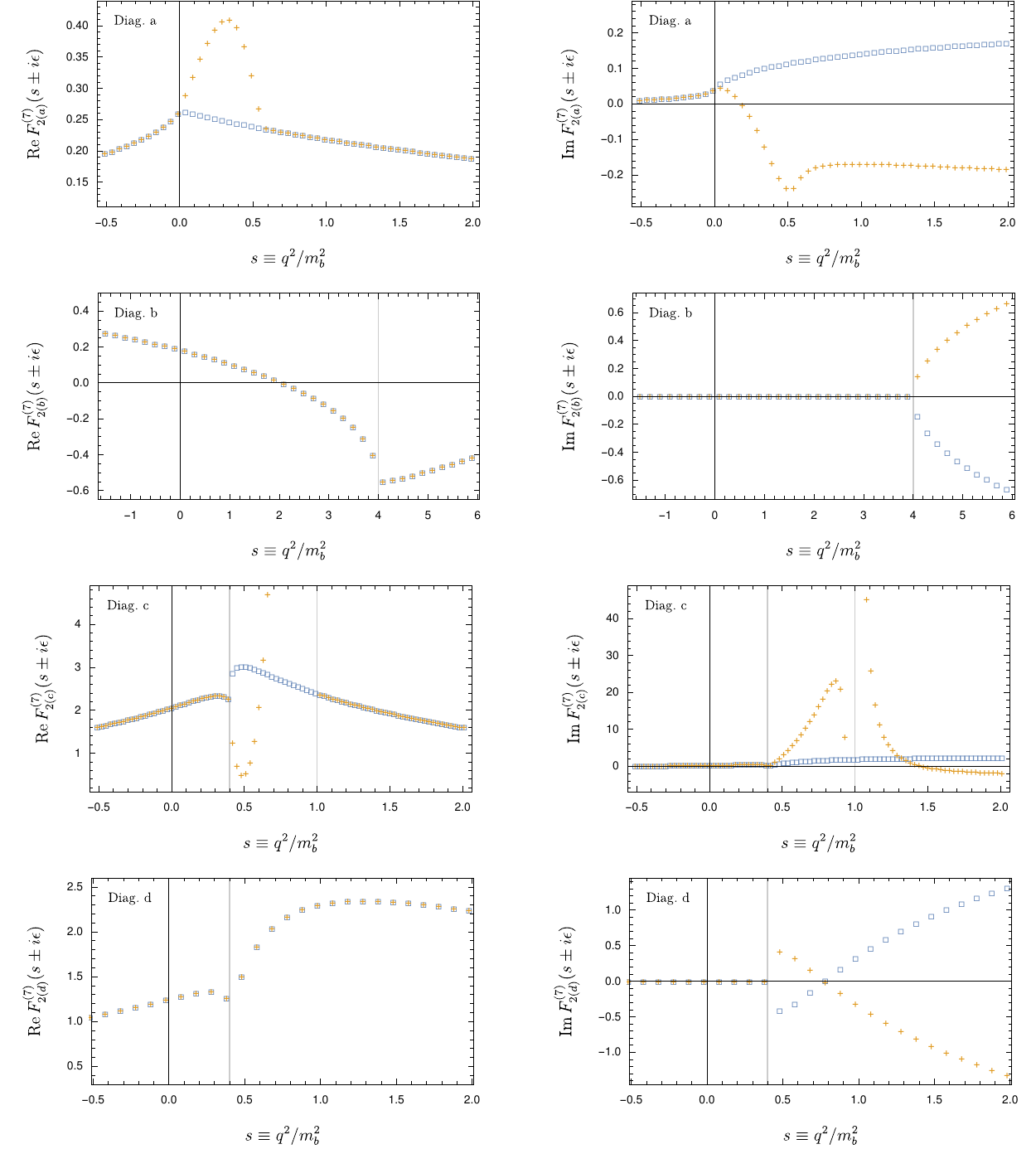}
\caption{Contributions to the form factor $F_{2}^{(7)}$ from each diagram class, evaluated above (blue squares) and below (orange crosses) the real axis. The discontinuities appear where expected and are real or imaginary as expected in each case. We have set $z=0.1$ and $\epsilon=10^{-8}$.}
\label{fig:discontinuities27}
\end{figure}

\begin{figure}
\includegraphics[width=\textwidth]{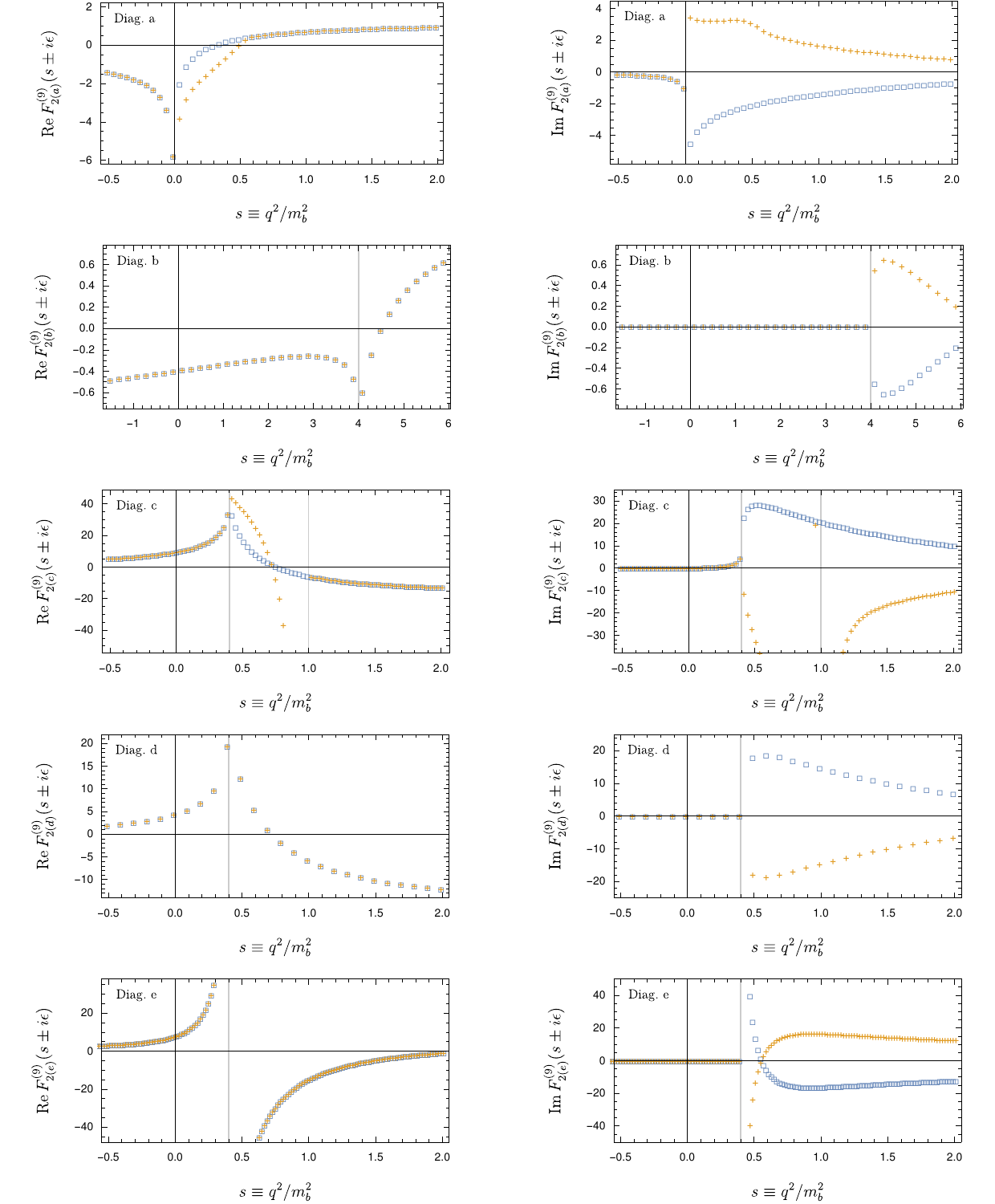}
\caption{Contributions to the form factor $F_{2}^{(9)}$ from each diagram class, evaluated above (blue squares) and below (orange crosses) the real axis. The discontinuities appear where expected and are real or imaginary as expected in each case. We have set $z=0.1$ and $\epsilon=10^{-8}$.}
\label{fig:discontinuities29}
\end{figure}

Concerning the discontinuities along the real axis, \Fig{fig:discontinuities27} and \Fig{fig:discontinuities29} show the contribution to the form factors for each diagram class, evaluated above and below the real axis, for a reference value of $z=0.1$. We see that the results obey the branch cut structure laid out above. Since the contributions from diagrams $b$, $d$ and $e$ are real below threshold, the branch-cut discontinuity is purely imaginary, as can be seen from the plots. On the contrary, the contributions from diagrams $a$ and $c$ are complex-valued below the thresholds since they have on-shell cuts in the variable $p_b^2$. This leads to a complex-valued branch-cut discontinuity (with a non-zero real part) in the ranges $0<s<4z$ and $4z<s<1$ respectively.

Besides explicitly confirming the expected branch-cut structure of the two-loop contributions, we find two features that we consider noteworthy:
\begin{itemize}

\item The discontinuities in diagrams $a$ and $c$ become purely imaginary for $s>4z$ and $s>1$, respectively.

\item The contribution from diagrams $c$ features a pole on the real axis when approaching the point $s=1$ from the negative imaginary plane. This pole is related to an anomalous threshold.
    
\end{itemize}

The same structure of branch cuts is found for the various counterterms: discontinuities starting at $s>0$, $s>4z$ and $s>1$ for $F_{i,Q_s}^{\text{ct}(7,9)}(s)$, $F_{i,Q_c}^{\text{ct}(7,9)}(s)$ and $F_{i,Q_b}^{\text{ct}(7,9)}(s)$ respectively. We refrain from showing the corresponding plots for brevity.

Concerning the dispersion relation, we have checked that~\Eq{eq:DR} is satisfied with good numerical accuracy separately for all diagram classes, each with its corresponding threshold. To give an example, we consider $F^{(7)}_{2,(b)}(s)$ with $z=0.1$. As discussed above, this function contains a branch cut starting at $s_{\rm th} = 4$. We find that its discontinuity can be fitted approximately by
\eqa{
{\rm Disc}\, F^{(7)}_{2,(b)}(s) &=& F^{(7)}_{2,(b)}(s+i0)-F^{(7)}_{2,(b)}(s-i0) 
\nonumber\\
&\simeq& i\, \theta(s-4)\,\bigg\{
-3.087 + e^{-0.0217\,s}
\bigg[
\frac{22.65}{s^2} - \frac{2.231}{s} +2.227
\nonumber\\
&& \hspace{22mm}+ 0.0532 s - 5.67\cdot 10^{-5}\,s^2 - 0.6028 \sqrt{s-4}
\bigg]
\bigg\}\ .
\label{eq:approxDisc}
}
Using this fit (for the sake of rapid integration) we find, for example taking $s_1= -3+i$ and $s_0=-1-2i$ in~\Eq{eq:DR}:
\eqa{
F^{(7)}_{2,(b)}(-3+i) - F^{(7)}_{2,(b)}(-1-2i) = 0.0894864 - 0.160827\,i\ ,
\\
\frac{-2+3i}{2\pi i} \int_{4}^\infty dt\  \frac{{\rm Disc}\, F^{(7)}_{2,(b)}(t)}{(t+3-i)(t+1+2i)} = 0.0894966 - 0.160839\,i\ .
}
As another example including a point at $s_0>0$: For $s_1= -1$ and $s_0=0.7$, we find:
\eqa{
F^{(7)}_{2,(b)}(-1) - F^{(7)}_{2,(b)}(0.7) = 0.117263\ ,
\\
\frac{-1.7}{2\pi i} \int_{4}^\infty dt\  \frac{{\rm Disc}\, F^{(7)}_{2,(b)}(t)}{(t+1)(t-0.7)} = 0.117265\ ,
}
again showing that the dispersion relation is very well verified. For applications with $s_0$ on the cut, the dispersion integral must include the prescription $(t-s_0-i\epsilon)$ in the denominator of the integrand, in order to regulate the pole (c.f.~\Eq{eq:disprel}). Thus, numerically the value taken for $\epsilon$ will determine the precision with which the discontinuity and the dispersion integral are evaluated.

\subsection{OPE coefficients with flavor separation}
\label{sec:FlavSep}

At this point we can collect the separate contributions to the OPE coefficients $\Delta C_{7,9}(q^2)$ proportional to the charge factors $Q_c$ and $Q_{s/b}$.  Denoting these two contributions by $\Delta C^{(c)}_{7,9}$ and $\Delta C^{(sb)}_{7,9}$,  they are given by
\eqa{
\Delta C^{(c)}_{7} &=& - \frac{\alpha_s}{4\pi} \sum_{i=1,2} C_i\,
\Big[ F_{i(c)}^{(7)} + F_{i(d)}^{(7)} + F_{i,Q_c}^{\text{ct}(7)} \Big]
\ ,
\label{eq:DeltaC7c}\\
\Delta C^{(c)}_{9} &=& f_{\rm LO}^{(9)} - \frac{\alpha_s}{4\pi} \sum_{i=1,2} C_i\,
\Big[ F_{i(c)}^{(9)} + F_{i(d)}^{(9)}  + F_{i(e)}^{(9)} + F_{i,Q_c}^{\text{ct}(9)} \Big]
\ ,\\[3mm]
\Delta C^{(sb)}_{7} &=& - \frac{\alpha_s}{4\pi} \sum_{i=1,2} C_i\,
\Big[ F_{i(a)}^{(7)} + F_{i(b)}^{(7)} + F_{i,Q_s}^{\text{ct}(7)} + F_{i,Q_b}^{\text{ct}(7)} \Big]
\ ,\\
\Delta C^{(sb)}_{9} &=& - \frac{\alpha_s}{4\pi} \sum_{i=1,2} C_i\,
\Big[ F_{i(a)}^{(9)} + F_{i(b)}^{(9)} + F_{i,Q_s}^{\text{ct}(9)} + F_{i,Q_b}^{\text{ct}(9)}
\Big]\ ,
}
where in~(\ref{eq:DeltaC7c}) we have omitted the term $F_{i,(e)}^{(7)}=0$.
These OPE coefficients will contribute separately to the functions $\H^{\rm OPE}_{\lambda,c}(q^2)$ and $\H^{\rm OPE}_{\lambda,sb}(q^2)$ appearing in the two different dispersion relations in~\Eq{eq:3DRs}. As discussed above, they have the proper analytic structure with branch cut discontinuities starting at $s>0$ and $s>4z$, for $\Delta C^{(sb)}_{7,9}$ and $\Delta C^{(c)}_{7,9}$ respectively.

\begin{figure}
\includegraphics[width=\textwidth]{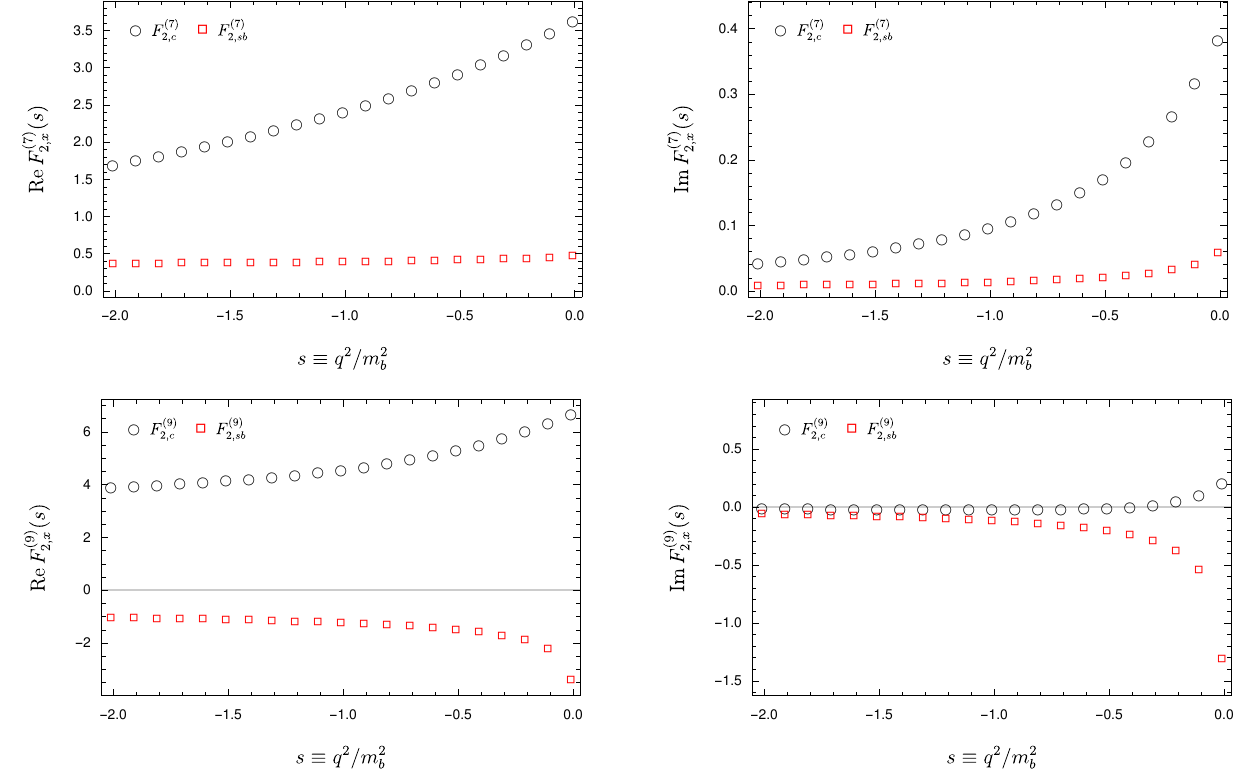}
\caption{Comparison between the two contributions proportional to $Q_c$ and $Q_{s/b}$ to the full renormalized form factors $F_{2}^{(7,9)}$, in the $q^2<0$ region. In these plots we have set $z=(0.29)^2$.}
\label{fig:plotFsbFc}
\end{figure}

A comparison of the size of the two different contributions to each NLO function is shown in~\Fig{fig:plotFsbFc}, where we plot the two functions $F_{2,c}^{(j)}$ and $F_{2,sb}^{(j)}$, defined by:
\eqa{
F_{i,c}^{(j)} &=& F_{i(a)}^{(j)} + F_{i(b)}^{(j)} + F_{i,Q_s}^{\text{ct}(j)} + F_{i,Q_b}^{\text{ct}(j)}\ ,
\\
F_{i,sb}^{(j)} &=& F_{i(c)}^{(j)} + F_{i(d)}^{(j)}  + F_{i(e)}^{(j)} + F_{i,Q_c}^{\text{ct}(j)}\ . 
}
The corresponding results for $F_{1,x}^{(j)}$ are qualitatively similar.
The conclusion is that, within the LCOPE region $q^2<0$,  the contribution proportional to the charge factor $Q_c$ is in most cases a few times larger than the one proportional to $Q_{s/b}$.

%%%%%%%%%%%%%%%%%%%%%%%%%%%%%%%%%%%%%%%%%%%%%%%%%
\section{Conclusions and outlook}
\label{sec:conclusions}
\setcounter{equation}{0}
%%%%%%%%%%%%%%%%%%%%%%%%%%%%%%%%%%%%%%%%%%%%%%%%

The determination of non-local effects in exclusive $b\to s\ell\ell$ processes is of great phenomenological interest, but very challenging theoretically. These effects are associated with the matrix element of a bi-local operator (c.f.~\Eq{eq:H}), which is significantly more complex than the usual ``local'' form factors that govern the naively-factorizable part of the amplitudes (such as the ones arising from semileptonic and electromagnetic dipole operators). The current approach to non-local effects is to write an OPE for the bi-local operator in a kinematic region where the OPE converges (even if unphysical) and then to extrapolate the results to the physical region using analyticity or dispersion relations. At the level of the OPE, the non-local matrix element can then be expressed in terms of simpler form factors, and OPE coefficients that are determined from a perturbative matching calculation.

The leading OPE coefficients have been known up to NLO for some time, but only in certain expansions on $q^2$ and/or $z=m_c^2/m_b^2$~\cite{Asatryan:2001zw,Greub:2008cy}. Here we have presented a recalculation of these two-loop contributions, fully analytic in both variables. This calculation has made use of the formalism of differential equations in canonical form, and the results are expressed in terms of Generalized Polylogarithms up to weight four. A particular attention has been put in obtaining an analytic continuation of the Feynman integrals with the desired singularity structure; for this purpose, special care is needed in fixing the integration constants in the solution of the differential equations.
Numerically, our results agree with previously known expanded results within their range of applicability, but deviate notably for $q^2 \lesssim -10\GeV^2$.

With the fully analytic results at hand, we have been able study the analytic properties of the non-local form factors, and we have confirmed the expectations from unitarity. In particular, we have verified the dispersion relations and checked the absence of singularities beyond the branch cuts from intermediate states in the $q^2$ channel.

In addition, we have presented the complete set of results separated into contributions proportional to different charge factors. This allows to study the extrapolation to the physical region separately for $c\bar c$ states, $s\bar s$ and $b\bar b$ states, and light states~\cite{Khodjamirian:2012rm,Bobeth:2017vxj}.

While the contributions from the operators $\cO_{1,2}$ considered here are the dominant ones in the SM for $b\to s$ transitions, it would be interesting to complete this calculation including the full set of four-quark operators in the general Weak Effective Theory~\cite{Aebischer:2017gaw}. This is important for an improved analysis beyond the SM~\cite{Jager:2019bgk}, and also for the case of $b\to d$ transitions, where the up-quark contributions are not CKM suppressed~\cite{Hambrock:2015wka}.

%%%%%%%%%%%%%%%%%%%%%%%%%%%%%%%%%%%%%%
\section*{Acknowledgements}

J.V. is grateful to Tobias Huber, Alex Khodjamirian, Bernhard Mistlberger, Jacobo Ruiz de Elvira and Danny van Dyk for useful discussions.
C.G. would like to thank J. Gasser for useful discussions and working out illuminating examples on dispersion relations and anomalous thresholds. 
H.M.A. is  supported by the Committee of Science of Armenia Program Grant No. 18T-1C162.
J.V. acknowledges funding from the European Union's Horizon 2020 research and innovation programme under the Marie Sklodowska-Curie grant agreement No 700525, `NIOBE' and from the Spanish MINECO through the “Ramon y Cajal” program RYC-2017-21870.
The work of C.G. is partially supported by the Swiss National Science Foundation under grant 200020-175449/1.

\newpage

\appendix
\renewcommand{\theequation}{\Alph{section}.\arabic{equation}} 

\setcounter{equation}{0}

\section{Details on ancillary files}

\subsection{A code to evaluate GPLs}
\label{app:GPLs}

As discussed in~\Sec{sec:NumGPLs}, we use GiNaC~\cite{ginac} and a C\texttt{++}-\texttt{Mathematica} interface to evaluate the GPLs appearing in our NLO results, and we provide this interface as an ancillary package here. The package includes two files:

\begin{enumerate}

\item The C\texttt{++} program \texttt{\color{blue} GPLs.cpp}. This program must be compiled and an executable with the name \texttt{GPLs.out} must be created. A typical command-line compilation would be
\begin{flushleft}
\texttt{g++ -std=c++11 GPLs.cpp -o GPLs.out -w  -lcln -lginac}
\end{flushleft}
where the appropriate libraries have been linked.
On Ubuntu, these libraries can be installed using the system package manager, e.g. via
\begin{flushleft}
\texttt{sudo aptitude install libginac-dev}
\end{flushleft}
The executable \texttt{GPLs.out}
uses GiNaC to evaluate GPLs with unit argument and no trailing zeroes (see~\Sec{sec:NumGPLs}).

\item The \texttt{Mathematica} program \texttt{\color{blue} GPLs.m}. This program defines the \texttt{Mathematica} routine
\begin{flushleft}
$\blacktriangleright\ $ \texttt{GPL[\{{\it weights}\},{\it argument}]}
\end{flushleft}
which expresses the original GPL in terms of GPLs with unit argument and no trailing zeroes, using~Eqs.(\ref{eq:4.2},\ref{eq:4.3}), and then uses \texttt{GPLs.out} to evaluate such GPLs.

\end{enumerate}

\subsection{Results for the functions $F_{1,2}^{(7,9)}$ in electronic form}
\label{app:F279}

The results for the renormalized two-loop functions $F_{1,2}^{(7,9)}$, as well as the separate contributions from each diagram class $F_{i({\rm diag})}^{(j)}$, ${\rm diag}=\{a,b,c,d,e\}$, and the counterterm contributions $F_{i,Q_q}^{{\rm ct}(j)}$ and $F_{i}^{{\rm ct}(j)}$, are given as well in \texttt{Mathematica} format as ancillary material.
We provide two \texttt{Mathematica} files:
\begin{enumerate}

\item The file \texttt{\color{blue} functionsNLO.m}. This program contains all the relevant LO and NLO functions:

\begin{itemize}

\item The LO functions \texttt{F170}, \texttt{F270}, \texttt{F190} and \texttt{F290} defined by
\eq{
f^{(7)}_{\rm LO} = C_1\, \texttt{F170} + C_2\, \texttt{F270} \ ,
\quad
f^{(9)}_{\rm LO} = C_1\, \texttt{F190} + C_2\, \texttt{F290}\ .
\nonumber
}
We note that  $\texttt{F170}=\texttt{F170}=0$.

\item The counterterm contributions:
\eq{
\texttt{F17ct} = F_{1}^{{\rm ct}(7)}\ ,
\quad
\texttt{F27ct} = F_{2}^{{\rm ct}(7)}\ ,
\quad
\texttt{F19ct} = F_{1}^{{\rm ct}(9)}\ ,
\quad
\texttt{F29ct} = F_{2}^{{\rm ct}(9)}\ ,
\nonumber
}
as well as the separate contributions with different charge factors,
\eqa{
\texttt{F17ctQs} = F_{1,Q_s}^{{\rm ct}(7)}\ ,
\quad
\texttt{F27ctQs} = F_{2,Q_s}^{{\rm ct}(7)}\ ,
\quad
\texttt{F19ctQs} = F_{1,Q_s}^{{\rm ct}(9)}\ ,
\quad
\texttt{F29ctQs} = F_{2,Q_s}^{{\rm ct}(9)}\ ,
\nonumber\\
\texttt{F17ctQc} = F_{1,Q_c}^{{\rm ct}(7)}\ ,
\quad
\texttt{F27ctQc} = F_{2,Q_c}^{{\rm ct}(7)}\ ,
\quad
\texttt{F19ctQc} = F_{1,Q_c}^{{\rm ct}(9)}\ ,
\quad
\texttt{F29ctQc} = F_{2,Q_c}^{{\rm ct}(9)}\ ,
\nonumber\\
\texttt{F17ctQb} = F_{1,Q_b}^{{\rm ct}(7)}\ ,
\quad
\texttt{F27ctQb} = F_{2,Q_b}^{{\rm ct}(7)}\ ,
\quad
\texttt{F19ctQb} = F_{1,Q_b}^{{\rm ct}(9)}\ ,
\quad
\texttt{F29ctQb} = F_{2,Q_b}^{{\rm ct}(9)}\ ,
\nonumber
}

\item The two-loop contributions from each diagram class:
\eqa{
&&
\texttt{F27b} = F_{2(b)}^{(7)}\ ,\quad
\texttt{F27c} = F_{2(c)}^{(7)}\ ,\quad
\texttt{F27d} = F_{2(d)}^{(7)}\ ,\quad
\texttt{F27e} = F_{2(e)}^{(7)}\ ,\quad
\nonumber\\
&&
\texttt{F29b} = F_{2(b)}^{(9)}\ ,\quad
\texttt{F29c} = F_{2(c)}^{(9)}\ ,\quad
\texttt{F29d} = F_{2(d)}^{(9)}\ ,\quad
\texttt{F29e} = F_{2(e)}^{(9)}\ ,
\nonumber
}
and \texttt{F27aupper}, \texttt{F29aupper}, \texttt{F27alower}, \texttt{F29alower} which correspond to $F_{2(a)}^{(7,9)}$ for positive and negative ${\rm Im}(s)$ respectively, as in this case the boundary conditions are fixed separately for the two cases (see~\Sec{sec:BoundaryConditions}).

\end{itemize}

All these functions are given in terms of the variables \texttt{xa}$=x_a$, \texttt{ya}$=y_a$, \dots , \texttt{xe}$=x_e$, \texttt{ye}$=y_e$ (c.f.~\Eq{eq:xiyi}), \texttt{vb}$=v_b$, \texttt{tb}$=t_b$ (c.f.~\Eq{eq:tbvb}),
\texttt{mub}$=\mu/m_b$, and the funcion \texttt{G} representing the GPL.

\item The program \texttt{\color{blue} FFNLO.m}. This is the master program to evaluate all the functions. It requires \texttt{GPL.m} and \texttt{functionsNLO.m} (which are evaluated at the beginning of the program), and defines two useful \texttt{Mathematica} routines:
\begin{flushleft}
$\blacktriangleright\ $ \texttt{FFNLO[$s,z,\mu/m_b$]}
\end{flushleft}
For given values of $s,z,\mu/m_b$ this routine calculates the full renormalized form factors $F_{1,2}^{(7,9)}$ (denoted by \texttt{F17}, \texttt{F27}, \texttt{F19} and \texttt{F29}), as well as the separate contributions discussed in~\Sec{sec:FlavSep}:
\eqa{
&&\texttt{F17Qc} = F_{1(c)}^{(7)} + F_{1(d)}^{(7)}  + F_{1(e)}^{(7)} + F_{1,Q_c}^{\text{ct}(7)} \ ,
\quad
\texttt{F17Qsb} = F_{1(a)}^{(7)} + F_{1(b)}^{(7)} + F_{1,Q_s}^{\text{ct}(7)} + F_{1,Q_b}^{\text{ct}(7)} \ ,
\nonumber\\
&&\texttt{F27Qc} = F_{2(c)}^{(7)} + F_{2(d)}^{(7)}  + F_{2(e)}^{(7)} + F_{2,Q_c}^{\text{ct}(7)}\ ,
\quad
\texttt{F27Qsb} = F_{2(a)}^{(7)} + F_{2(b)}^{(7)} + F_{2,Q_s}^{\text{ct}(7)} + F_{2,Q_b}^{\text{ct}(7)}\ ,
\nonumber\\
&&\texttt{F19Qc} = F_{1(c)}^{(9)} + F_{1(d)}^{(9)}  + F_{1(e)}^{(9)} + F_{1,Q_c}^{\text{ct}(9)}\ ,
\quad
\texttt{F19Qsb} = F_{1(a)}^{(9)} + F_{1(b)}^{(9)} + F_{1,Q_s}^{\text{ct}(9)} + F_{1,Q_b}^{\text{ct}(9)}\ ,
\nonumber\\
&&\texttt{F29Qc} = F_{2(c)}^{(9)} + F_{2(d)}^{(9)}  + F_{2(e)}^{(9)} + F_{2,Q_c}^{\text{ct}(9)}\ ,
\quad
\texttt{F29Qsb} = F_{2(a)}^{(9)} + F_{2(b)}^{(9)} + F_{2,Q_s}^{\text{ct}(9)} + F_{2,Q_b}^{\text{ct}(9)}\ ,
\nonumber
}
and gives as a result a replacement rule for all twelve functions. 
\begin{flushleft}
$\blacktriangleright\ $ \texttt{FFapplied[$s,z,\mu/m_b$,{\it function}]}
\end{flushleft}
For given values of $s,z,\mu/m_b$, this routine evaluates the function {\it function}, which can be any of the functions defined in \texttt{functionsNLO.m} (thus allowing the evaluation of the individual contributions to $F_{1,2}^{(7,9)}$), or in fact any function involving \texttt{G} functions (GPLs).

These routines operate by first collecting a list of the different GPLs that appear, in order to evaluate each GPL only once. This leads to a huge increase in the speed of the evaluation. 
    
\end{enumerate}

\section{List of Master Integrals}
\label{app:MIs}

In this appendix we collect the list of all Master Integrals (MIs) $J_{i,k}$ that appear in the calculation of the two-loop diagrams $a$-$e$ in~\Fig{fig:diags}.
The notation is described in~\Sec{sec:3.2}.

\bigskip

\noindent For diagrams $a$ there are 7 MIs:
\begin{align}
& J_{a,1}=j[a,1,1,0,0,0,0,0]
& J_{a,2}=j[a,1,1,0,0,1,0,0]
&& J_{a,3}=j[a,2,1,0,0,1,0,0]
\nonumber\\
& J_{a,4}=j[a,0,1,0,1,1,0,0]
& J_{a,5}=j[a,0,1,1,0,1,0,0]
&& J_{a,6}=j[a,1,1,0,1,1,0,0]
\\
& J_{a,7}=j[a,2,1,0,1,1,0,0]
\nonumber
\end{align}

\bigskip

\noindent For diagrams $b$ there are 9 MIs:
\begin{align}
& J_{b,1}=j[b,0,1,0,0,1,0,0] 
& J_{b,2}=j[b,1,1,0,0,0,0,0]
&& J_{b,3}=j[b,1,1,0,1,0,0,0]
\nonumber\\
& J_{b,4}=j[b,1,1,0,0,1,0,0]
& J_{b,5}=j[b,2,1,0,0,1,0,0]                   
&& J_{b,6}=j[b,0,1,0,1,1,0,0]
\\ 
& J_{b,7}=j[b,1,1,0,1,1,0,0]
& J_{b,8}=j[b,2,1,0,1,1,0,0]
&& J_{b,9}=j[b,1,1,0,2,1,0,0]
\nonumber
\end{align}

\bigskip

\noindent For diagrams $c$ there are 9 MIs:
\begin{align}
& J_{c,1}=j[c,0,1,1,0,0,0,0]  
& J_{c,2}=j[c,1,0,1,0,1,0,0]
&& J_{c,3}=j[c,1,0,1,1,0,0,0]
\nonumber\\
& J_{c,4}=j[c,1,1,1,0,0,0,0]
& J_{c,5}=j[c,1,1,1,0,1,0,0]                    
&& J_{c,6}=j[c,1,2,1,0,1,0,0]
\\ 
& J_{c,7}=j[c,2,0,1,0,1,0,0]
& J_{c,8}=j[c,2,0,1,1,0,0,0]
&& J_{c,9}=j[c,2,1,1,0,1,0,0]
\nonumber
\end{align}

\bigskip

\noindent For diagrams $d$ there are 15 MIs:
\begin{align}
& J_{d,1}=j[d,0,1,1,0,0,0,0]  
& J_{d,2}=j[d,0,0,1,0,1,0,0]
&& J_{d,3}=j[d,0,1,1,0,1,0,0]
\nonumber\\
& J_{d,4}=j[d,0,1,1,1,0,0,0]
& J_{d,5}=j[d,0,2,1,1,0,0,0]
&& J_{d,6}=j[d,1,0,1,0,1,0,0]
\nonumber\\
& J_{d,7}=j[d,2,0,1,0,1,0,0]  
& J_{d,8}=j[d,1,1,0,0,1,0,0]
&& J_{d,9}=j[d,1,1,1,0,0,0,0]
\\
& J_{d,10}=j[d,0,1,1,1,1,0,0]
& J_{d,11}=j[d,0,2,1,1,1,0,0]                      
&& J_{d,12}=j[d,1,1,1,0,1,0,0]
\nonumber\\ 
& J_{d,13}=j[d,1,2,1,0,1,0,0]
& J_{d,14}=j[d,2,1,1,0,1,0,0]
&& J_{d,15}=j[d,1,1,2,0,1,0,0]
\nonumber
\end{align}

\bigskip

\noindent For diagrams $e$ there are 5 MIs:
\begin{align}
& J_{e,1}=j[e,0,1,0,0,1,0,0]
& J_{e,2}=j[e,0,1,0,1,1,0,0]
&& J_{e,3}=j[e,0,1,1,0,1,0,0]
\\
& J_{e,4}=j[e,0,2,0,1,1,0,0]
& J_{e,5}=j[e,1,1,1,0,1,0,0]
\nonumber
\end{align}

\section{Weights}
\label{app:weights}

In this appendix we collect the different weights appearing in the GPLs.
In GPLs with argument $x_i$, the weights are constants:
\eq{
w_0 = 0\ ,\quad
w_1 = 1\ ,\quad
w_2 = i\ ,\quad
w_3 = 2+\sqrt3\ ,\quad
w_4 = 2-\sqrt3\ . 
}
In GPLs with argument $y_i$, $v_b$ or $t_b$, the weights are $x_i$-dependent (with $x_i$ depending on the diagram class):
\eqa{
&&w_0(x) = 0 \ ,\quad
w_1(x) = 1 \ ,\quad
w_2(x) = x \ ,\quad
w_3(x) = 2x^2/(1+x^2) \ ,\quad
w_4(x) = 2x/(1-x)^2\ ,
\nonumber\\[1mm]
&&w_5(x) = 2x/(1+x)^2 \ ,\quad
w_6(x) = 2ix/(1-x^2) \ ,\quad
w_7(x) = 8x^2/(1-6x^2+x^4)\ ,
\nonumber\\[1mm]
&&w_{10}(x) = (4x^2 - 2\sqrt2\sqrt{x^2 + 4 x^4 + x^6})/(1 + x^2)^2\ ,  \\[1mm]
&&w_{11}(x) = (4x^2 + 2\sqrt2\sqrt{x^2 + 4 x^4 + x^6})/(1 + x^2)^2\ .
\nonumber
}

\section{Explicit examples for fixing integration constants}
\label{app:ExamplesBCs}

We first consider the master integral from diagram $e$ with four propagators, i.e. $J_{e,5}$. Solving the corresponding differential equations in the canonical basis and then transforming the solution to the ordinary basis we get, for the $\epsilon^{-2}$ part of  $J_{e,5}$
\begin{equation} 
J_{e,5}^{(-2)}=-\frac{1}{256 \pi^4} + \frac{9 c_2 + \frac{1}{256 \pi^4}}{y_e^2} \, ,
\end{equation}
where $c_2$ is an integration constant.
Imposing the condition that $J_{e,5}^{(-2)}$ is nonsingular for $s\to 0$ (which is equivalent  to $y_e \to 0$), we get  $c_2=-\frac{1}{2304 \pi^4}$, leading to
\begin{equation} 
J_{e,5}^{(-2)}=-\frac{1}{256 \pi^4} \, . 
\end{equation}
In the same way we get, for the $\epsilon^{-1}$ part of $J_{e,5}$,
\begin{eqnarray}
\nonumber J_{e,5}^{(-1)} &=&\frac{1}{128 \pi^4 y_e^2}\left [1 +i \pi + 1152 c_1 \pi^4 - 2 y_e^2 
- i \pi y_e^2 +    y_e^2 G(-1; x_e) +y_e^2 G(1; x_e) \right.
\\ && \left. - 2 y_e^2 G(0; x_e) + y_e G(-1; y_e) - y_e G(1; y_e) + 
   2 \log(2) - 2y_e^2 \log(2)\right].
\end{eqnarray}
Again imposing the condition that $J_{e,5}^{(-1)}$ is nonsingular for $y_e\to 0$,
we obtain $c_1= -\frac{1 + i \pi + 2 \log(2)}{1152 \pi^4}$, leading to
\begin{eqnarray}
\nonumber J_{e,5}^{(-1)} &=&\frac{1}{128 \pi^4 y_e}\left [ - 2 y_e 
- i \pi y_e +    y_e G(-1; x_e) +y_e G(1; x_e) \right.
\\ && \left. - 2 y_e G(0; x_e) +  G(-1; y_e) - G(1; y_e)  - 2y_e \log(2)\right] \, .
\end{eqnarray}
The results for  $J_{e,5}^{(0)}$ and $J_{e,5}^{(1)}$ are obtained analogously.

As a second example we consider the MIs $J_{c,2}$  and $J_{c,7}$ of diagram $c$.
Solving the corresponding differential equations in the canonical basis and then transforming the solution to the ordinary basis, we get for the  $\epsilon^{-2}$ parts of  $J_{c,2}$ and  $J_{c,7}$,
\begin{eqnarray}
\nonumber J_{c,2}^{(-2)}&=&\frac{(x_c-1) (x_c+1)
   \left(12288 \pi ^4 c_2^1
   x_c^2+4096 \pi ^4
   c_2^1-x_c^2+1\right)}{
   4096 \pi ^4 x_c^4} \, , \\
   J_{c,7}^{(-2)}&=&\frac{4096 \pi ^4 c_2^1
   x_c^2+4096 \pi ^4
   c_2^1-x_c^2+1}{1024
   \pi ^4 x_c^2} \, .
\end{eqnarray}
$J_{c,2}$ has three propagators. $J_{c,7}$ also has three propagators but
one of them is squared. This means that $J_{c,2}\sim z$ and
$J_{c,7}\sim z^0$ for large $z$. Or in terms of $x_c$, $J_{c,2}\sim x_c^{-2}$ and $J_{c,7}\sim x_c^0$ when $x_c \to 0$. Imposing these conditions, we find
$c_2^1=-\frac{1}{4096 \pi ^4}$, leading to
 \begin{eqnarray}
\nonumber J_{c,2}^{(-2)}&=&\frac{(1-x_c)
   (1+x_c)}{1024 \pi ^4
   x_c^2} \, , \\
   J_{c,7}^{(-2)}&=&-\frac{1}{512 \pi ^4} \, .
\end{eqnarray}
In the same way one can derive the results for  $J_{c,2}^{(-1)},J_{c,7}^{(-1)}$, 
$J_{c,2}^{(0)},J_{c,7}^{(0)}$, and $J_{c,2}^{(1)},J_{c,7}^{(1)}$.

 \newpage

%%%%%%%%%%%%%%%%%%%%%%%%%%%%%%%%%%%%%%%%%%%%%%%%%%%%%%%%%%%%%%%%%%%%%%%%%%%%%%%%%%%%

\end{document}